\documentclass[aps,prl,twocolumn]{revtex4-1}
\usepackage{graphicx}
\usepackage{color}
\usepackage{xcolor}

\usepackage{lineno}
\begin{document}


\title{Metallization of shock-compressed liquid ammonia}


\author{A.~Ravasio$^1$}\email[]{alessandra.ravasio@polytechnique.fr}
\author{M.~Bethkenhagen$^{2,3}$ }
\author{J.-A.~Hernandez$^{1,4}$}
\author{A.~Benuzzi-Mounaix$^1$}
\author{F.~Datchi$^5$}
\author{M.~French$^3$}
\author{M.~Guarguaglini$^1$} 
\author{F.~Lefevre$^1$}
\author{S.~Ninet$^5$}
\author{R.~Redmer$^3$}
\author{T.~Vinci$^1$}

\affiliation{$^{1}$ LULI, CNRS, CEA, \'Ecole Polytechnique - Institut Polytechnique de Paris, route de Saclay, 91128 Palaiseau cedex, France \\
$^{2}$\'Ecole Normale Sup\'erieure de Lyon,  Universit\'e Lyon 1, Laboratoire de G\'eologie de Lyon, CNRS UMR 5276, 69364 Lyon Cedex 07, France\\
$^{3}$Institut f\"ur Physik, Universit\"at Rostock, 18051 Rostock, Germany\\
$^{4}$Centre for Earth Evolution and Dynamics, University of Oslo, Norway\\
$^{5}$ Institut de Min\'eralogie, de Physique des Mat\'eriaux et de Cosmochimie (IMPMC), Sorbonne Universit\'e, CNRS UMR 7590, IRD UMR 206, MNHN, 4 place Jussieu, F-75005 Paris, France}


\date{\today}

\begin{abstract}

Ammonia is predicted to be one of the major components in the depths of the ice giant planets Uranus and Neptune. Their dynamics, evolution, and interior structure are insufficiently understood and models 
rely imperatively on data for equation of state and transport properties. Despite its great significance, the experimentally accessed region of the ammonia phase diagram today is still very limited in 
pressure and temperature. Here we push the probed regime to unprecedented conditions, up to $\sim$350 GPa and $\sim$40 000 K. Along the Hugoniot, the temperature measured as a function of pressure shows a 
subtle change in slope at $\sim$7000 K and $\sim$90~GPa, in agreement with ab initio simulations we have performed. This feature coincides with the gradual transition from a molecular liquid to a plasma 
state. Additionally, we performed reflectivity measurements, providing the first experimental evidence of electronic conduction in high-pressure ammonia. Shock reflectance continuously rises with pressure 
above 50~GPa and reaches saturation values above 120~GPa. Corresponding electrical conductivity values are up to one order of magnitude higher than in water in the 100~GPa regime, with possible significant 
contributions of the predicted ammonia-rich layers to the generation of magnetic dynamos in ice giant interiors.
 
\end{abstract}


\maketitle

\begin{figure}[]
\centering
\includegraphics[width=\linewidth]{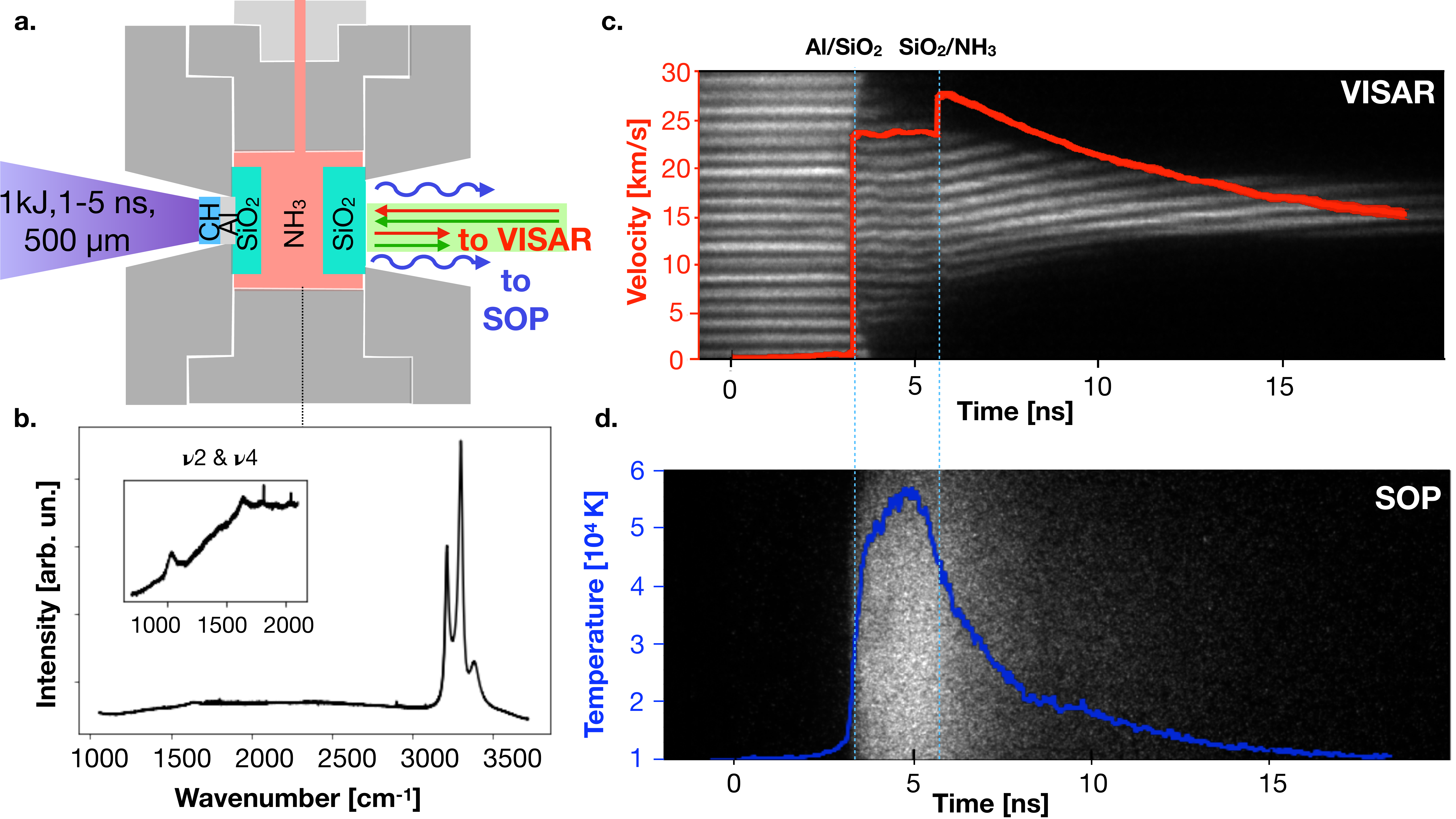}
\caption{a. Schematic experimental setup of the laser pulse inducing shock compression in the liquid ammonia sample. b. Raman spectrum of the sample indicative of pure liquid NH$_3$ with $\nu_1$-$\nu_3$ 
stretching band around 3300 cm$^{-1}$ as well as $\nu_2$ and $\nu_4$ modes (inset) at 1200 cm$^{-1}$ and 1600 cm$^{-1}$, respectively. c. VISAR signal  and d. SOP data together with the extracted velocity 
and temperature measurements. For this shot, laser energy and pulse length were respectively 816~J and 2~ns.}
\label{fig: setup}
\end{figure}

As an archetypal hydrogen-bonded system, the properties of ammonia (NH$_3$) at high pressures (P) and temperatures (T)  are of particular interest in solid state physics and chemistry \cite{Liu2001,Fortes2003,Hermann2017}. 
For temperatures up to 10~000~K and pressures up to 500~GPa, theoretical studies predict an exceptionally rich phase diagram, including fluid phases of different chemical composition, solid and superionic 
phases~\cite{Cavazzoni1999,Pickard2008,Bethkenhagen2013}. Unveiling these extremely intriguing behaviours has also crucial impact on planetary science because NH$_3$ has a significant cosmic abundance. As 
a part of the so called ``planetary ice" (mixtures of water, ammonia, and methane), it is believed to be found in a wide range of thermodynamic conditions in planets and their satellites \cite{Hubbard1980,Hubbard1991,Journaux2020,Helled2020}, 
both within our Solar System and beyond \cite{Borucki2011,Fressin2013,Zeng2019}. Adiabatic interior models of Uranus and Neptune predict that planetary ice exists between 20 to 600 GPa and 
2000 to 7000 K \cite{Hubbard1980,Redmer2011, Helled2011,Nettelmann2013,Bethkenhagen2017, Scheibe2019}, which possibly coincides with the stability range for superionic ammonia. 
Recent models \cite{Nettelmann2016, Podolak2019,Vazan2020} more consistent with Uranus’ low luminosity observations predict much warmer interiors, with temperatures up to 25 000~K, where ammonia would 
rather be a warm dense ionized fluid ~\cite{Cavazzoni1999, Bethkenhagen2013,Li2013,Li2017}. These non-adiabatic models highly rely on transport properties of the planetary ice at such high temperatures, 
which are mostly unavailable to date. Moreover, the discovery of hot Neptune exoplanets, like Gliese 436b, further extends the necessity of exploring a wide range of thermodynamic 
conditions \cite{Nettelmann2010}. Accurate determination of the high-pressure ammonia equation of state is hence required to build more reliable interior models accounting for planetary ice as a complex 
mixture~\cite{Nettelmann2016,Bethkenhagen2017,Guarguaglini2019} and overcoming the often used, yet too simplistic, approach that considers planetary ice as made of pure water.
Similarly, the NH$_3$ transport properties, such as electrical conductivity, are crucial to address Uranus' and Neptune's peculiar magnetic fields~\cite{Stevenson2010,Stanley2006,Soderlund2013}. \\
Despite its importance, experimental studies of the high pressure behavior of ammonia remain very scarce and limited to a narrow thermodynamic range. Measurements in diamond anvil cells have characterized 
the equation of state and phase diagram up to 200~GPa at 300~K, but reached at most $\sim$40~GPa at higher temperatures ($\sim$3000~K) \cite{Ninet2008,Li2009,Ninet2012,Ninet2014,Queyroux2019,Queyroux2019b,Ojwang2012,Palasyuk2014}. 
Few studies have been carried out employing dynamic compression methods, mainly at gas-gun facilities, providing shock compression data up to 70~GPa~\cite{Dick1981,Mitchell1982,Nellis1988,Radousky1990,Nellis1997,Dattelbaum2019}. 
However, no reflectivity data were obtained, only two temperature points were reported~\cite{Radousky1990} and only two electrical conductivity data points exceeded 40~GPa~\cite{Nellis1997}.
The dearth in experimental data is undoubtedly linked to the severe difficulties in preparing appropriate samples for high pressure experiments, in particular those using laser-driven shocks.\\
In this work, we report on the first laser-generated shock compression experiment in ammonia, which allowed us to study the equation of state and optical reflectivity (tightly linked to electronic 
conductivity) in an unprecedented regime, up to $\sim$ 350 GPa and 40~000 K, 
together with new \textit{ab initio} calculations.\\
The experiment was performed at the LULI2000 laser facility at the LULI Laboratory in France. The setup is shown schematically in Fig. \ref{fig: setup}a. A long pulse ($\tau_L$ $\sim$  1.5-5~ns) high-energy
($E_L$ up to 1~kJ at $\lambda_L=$ 527~nm) laser beam was focused ($\sim$ 500~$\mu$m smoothed focal spot) on the target to generate a compression shock wave. In order to reach relevant pressures, we have 
conceived and assembled a specific target capable to accommodate liquid ammonia and compatible with standard laser-shock experiments. Gaseous NH$_3$ was first condensed at low temperatures ($\sim$ 243~K) 
in a small stainless steel cell-target. Two thin $\alpha$-quartz windows of 50~$\mu$m and 250~$\mu$m were used at the front (laser-side) and back (diagnostic side) of the cell respectively, to ensure 
tightness and allow optical probing. A CH-Al pusher was glued on the outer part of the front side SiO$_2$ window to avoid any contamination of the sample. The system was then brought to ambient temperature,
allowing an outgoing controlled flow to maintain an almost constant pressure and avoid breaking of the thin quartz windows under the thermal pressure increase. Finally, the initial Hugoniot state was fixed 
at 14 bar, room temperature ($\sim$ 295 K), and a density of 0.61~g/cm$^{3}$ (based on NIST database). The Raman spectrum of the filled cell is consistent with that of pure liquid ammonia without any 
measurable impurities (see Fig. \ref{fig: setup}b).
Visible diagnostics were used to probe shocked ammonia. A dual-channel velocity interferometer system for any reflector (VISAR) \cite{Barker1972,Celliers2004_VISAR} operated at 532~nm and 1064~nm 
(sensitivities of 4.96~km/s and 12.81~km/s) was used to measure the time-dependent shock velocity for equation of state measurements and reflectivity. The shock-front reflectivity at 532~nm was also 
measured independently with a reflectometer. A streaked optical pyrometer (SOP) \cite{Holmes1995,Hall1997,Miller2007} collected the self-emission of the shocked sample as a function of time to gather 
temperature estimation. 
Typical experimental data are shown in Fig.\ref{fig: setup}c/d. Quartz (front side window) and ammonia are both rapidly ionized and transformed into a reflecting state upon shock compression. The 
instantaneous shock velocity in both materials could therefore be measured with the VISARs, knowing the values of the quartz and ammonia pristine refractive indices ($n_0^{Qz}$ and $n_0^{NH_3}$ respectively) 
(see Supplementary Material). 
Using quartz as \textit{in situ} standard and applying impedance mismatch \cite{forbes, Brygoo2015} together with the Rankine-Hugoniot relations \cite{zeldovich} we get the thermodynamic conditions (mass 
density $\varrho$, pressure $P$ and internal energy $E$) in shocked ammonia at the SiO$_2$/NH$_3$ interface \footnote{See Supplementary Material for further experimental details, which includes 
Refs.~\cite{Ghosh1999, Gauthier1988, Kume1998, Robertson1973, Tillner1993, Huser2015, Qi2015, Gregor2016}}.  

In addition to the experiment, we performed \textit{ab initio} simulations using density functional theory molecular dynamics (DFT-MD) using VASP~\cite{Kresse1993a,Kresse1993b,Kresse1994,Kresse1996}. The 
Hugoniot states were calculated based on previous \textit{ab initio} EOS data~\cite{Bethkenhagen2013} derived with the Perdew-Burke-Ernzerhof (PBE) exchange-correlation functional~\cite{Perdew1996}. Here, 
we extended these calculations and calculated reflectivity and DC electrical conductivity with both the PBE and the HSE~\cite{Heyd2003,Heyd2006} hybrid functional \footnote{See Supplementary Material for 
further computational details, which includes Refs.~\cite{Nose1984, Baldereschi1973, Kubo1957, Greenwood1958, Holst2011, Gajdos2006, Berens1983}}.


%
\begin{figure}[h]
\centering
\includegraphics[width=\linewidth]{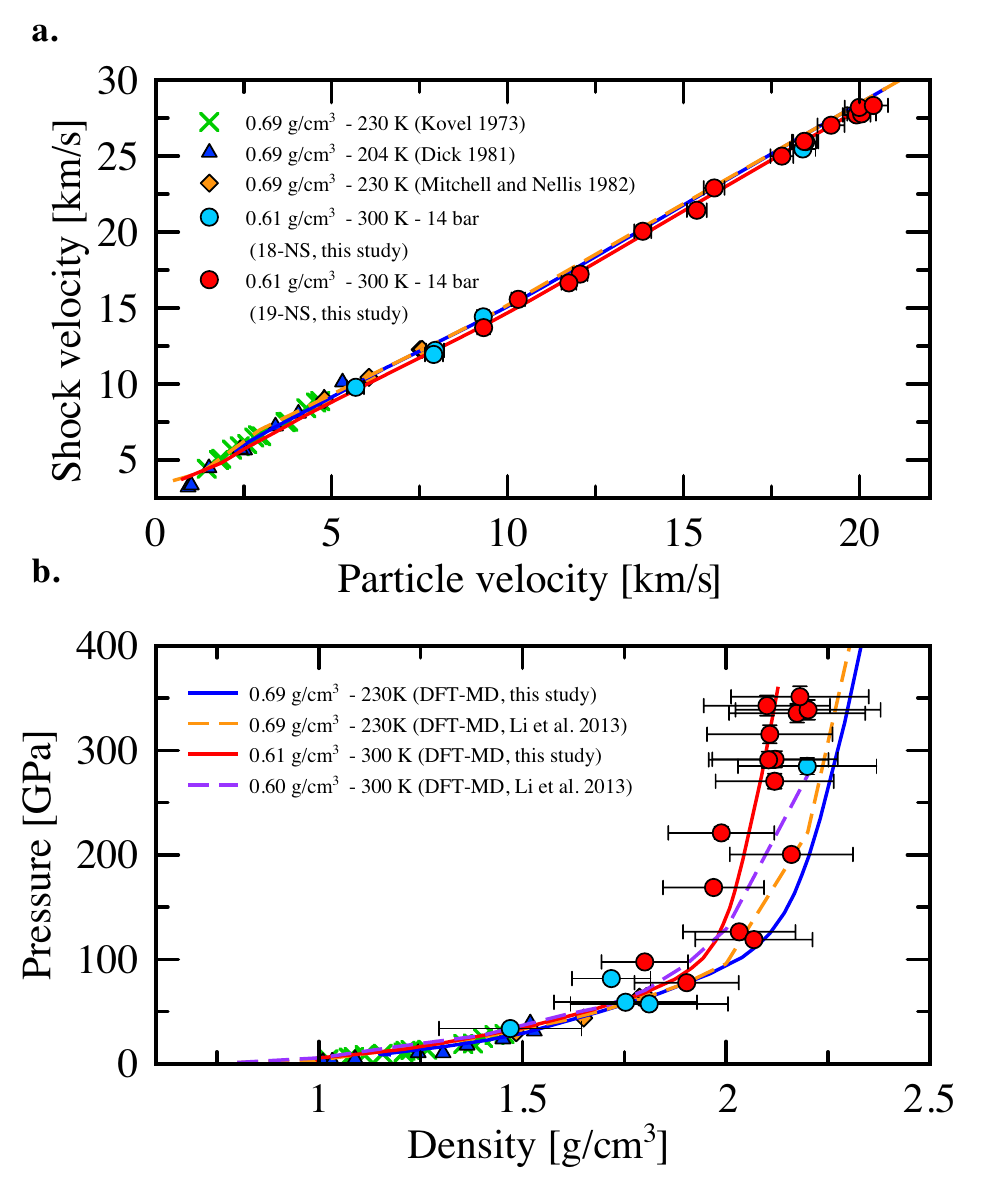}
\caption{Hydrodynamic conditions probed in shocked liquid ammonia ($\varrho_0$=0.61~g/cm$^{3}$) a. Relation between shock and particle velocities. 
Also shown are previous measurements at an initial density of 0.69~g/cm$^{3}$ \cite{Kovel1973,Dick1981,Mitchell1982} and the previous (dashed lines) and present (continuous lines) DFT-MD data considering 
both initial states. b. Corresponding pressure--density diagram.  
}
\label{fig: measurements}
\end{figure}
The experimental data for the Hugoniot are presented in Fig.~\ref{fig: measurements}. 
A total of 20 Hugoniot data points were recorded over the range 35 -- 350~GPa, represented as dot symbols in Fig.~\ref{fig: measurements}. Also shown are previous experimental results from gas-gun 
facilities (starting densities $\varrho_0=0.69$~g/cm$^{3}$ ) and the Hugoniots from \textit{ab initio} calculations performed in this work ($\varrho_0$=0.61~g/cm$^{3}$ and $\varrho_0$=0.69~g/cm$^{3}$) and 
in Li {\it et al.}~\cite{Li2013} ($\varrho_0=0.60$~g/cm$^{3}$ and $\varrho_0$=0.69~g/cm$^{3}$). For low pressures, the Hugoniot curves are rather insensitive to the slightly different initial densities and 
our data are consistent with the gas gun data. Moreover, our experimental data set is consistent with our theoretical predictions over the whole probed region.
Because of the large impedance mismatch and the consistent thickness of the target, the shock wave is not sustained during its propagation in the ammonia sample. Following the decay of the compression wave 
as it traveled through the sample, we also obtained continuous time measurements of the temperature and reflectivity as a function of shock velocity (or of pressure, density and internal energy, after 
applying the corresponding relation from the Hugoniot data). 

\begin{figure}[t]
\includegraphics[width=1.1\linewidth]{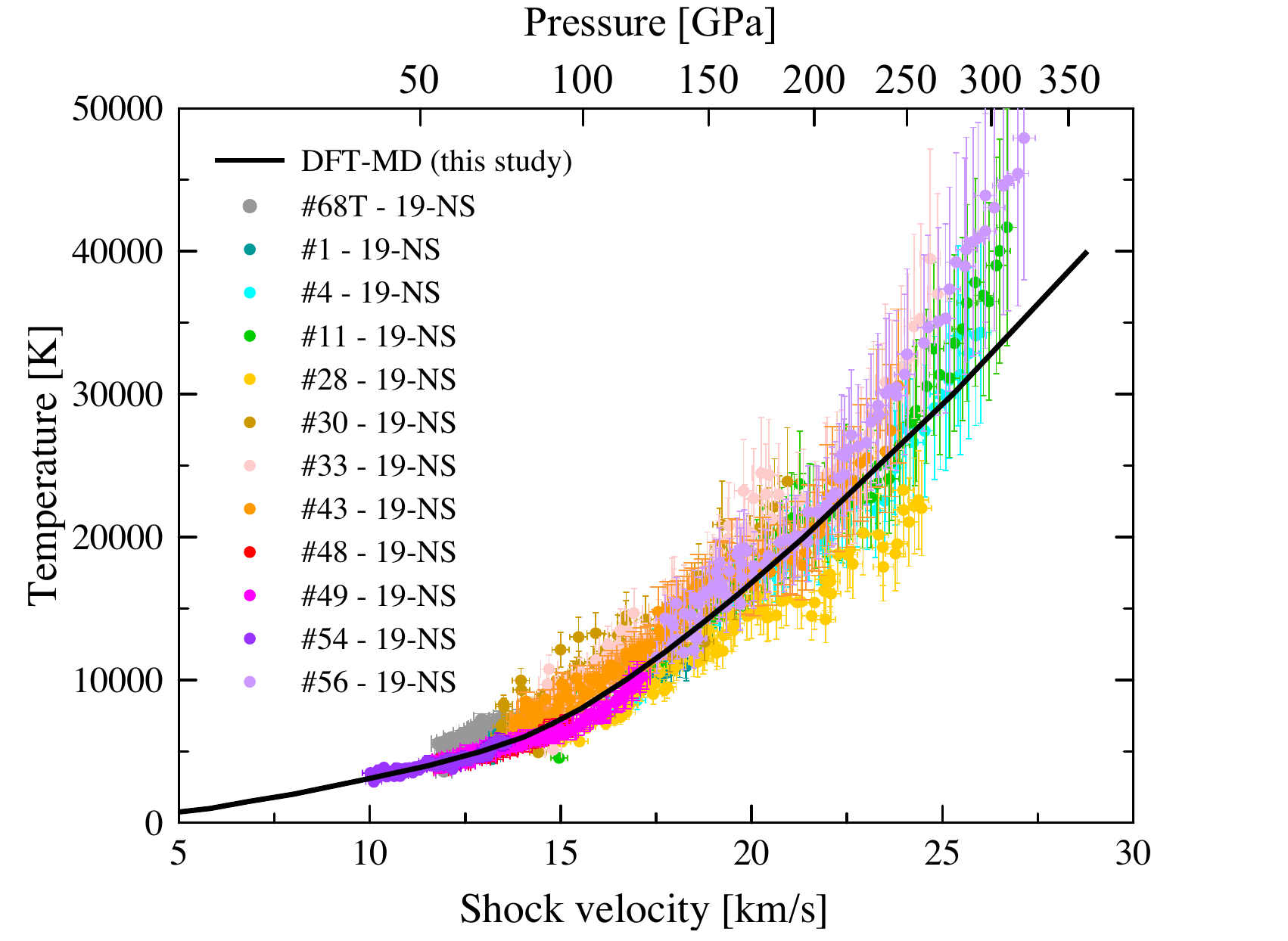}
\caption{Experimental (circles) and theoretical (black line) temperature of shocked ammonia as function of the shock velocity along the 0.61 g/cm$^{3}$  Hugoniot. Each circle represents a time-resolved 
measurement during the propagation of a decaying shock with typical error bars. The pressure scale on the top x-axis is obtained by applying the $U_{s}-U_{p}$ linear relation derived from experimental 
measurements.}
\label{fig: temperature}
\end{figure}
\begin{figure}[htb]
\includegraphics[width=0.86\linewidth]{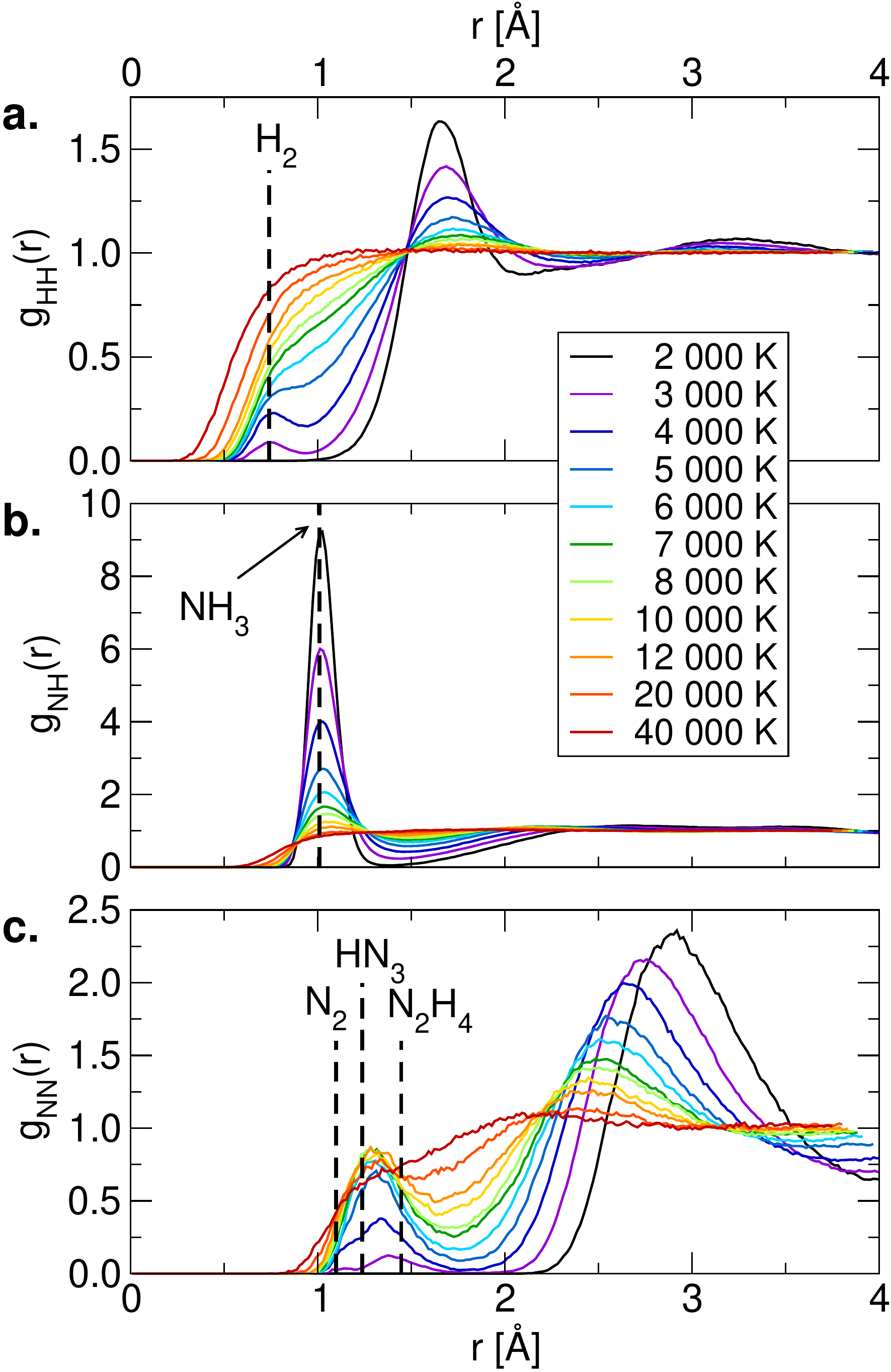}
\caption{a. H-H, b. N-H, and c. N-N pair correlation functions for different temperatures along the Hugoniot (solid lines). The dashed lines indicate bond lengths in various molecules that were visually 
observed in the simulations.}
\label{fig: calculations}
\end{figure}

Temperature is one of the most difficult parameters to access and up to now only two experimental data points were available at $\sim$4000~K \cite{Radousky1990}. Our data set, shown in 
Fig.~\ref{fig: temperature}, provides a much wider ensemble up to very high temperatures. The consistency with calculations is remarkable, except for the highest points where the experimental 
uncertainties are larger. In particular, in both calculations and experiment we find that the shock temperature increases continuously up to $\sim$ 7000~K ($\sim$ 90~GPa), where a slight change of 
slope occurs, see Fig.~\ref{fig: temperature}. In solids this kind of behavior is often linked to phase transitions, including melting (ex. \cite{Millot2015, McWilliams2012}). To understand the origin of 
this feature in ammonia, we have computed the site-site pair correlation functions along the Hugoniot, shown in Fig.~\ref{fig: calculations}. The analysis of these calculations indicates that the observed 
change in slope in the $P$-$T$ curve is caused by dissociation processes in the molecular liquid and its transformation into a plasma state. In the N-H pair correlation function in 
Fig.~\ref{fig: calculations}b, the drop of the intramolecular bond peak at 1.08~\AA~with increasing temperature illustrates the gradual dissociation of the ammonia molecules. We also observe additional 
peaks in the H-H and N-N pair correlation functions above 2000~K, which are indicative of the formation of H$_2$ and N$_2$ molecules. Even more complex and very short-lived molecules such as N$_2$H$_4$ 
and HN$_3$ were observed from visual inspections of the ionic trajectories in the DFT-MD. It can be seen that the characteristic H$_2$ signal in Fig.~\ref{fig: calculations}a vanishes above 5000~K and 
that the broad feature corresponding to different forms of N-N bonding (between 1.0~\AA~and 1.5~\AA~in Fig.~\ref{fig: calculations}c) reaches its maximum at 7000~K--8000~K. This implies that the molecular 
NH$_3$ liquid transforms through continuous dissociation and ionization processes into a plasma. Additionally, diffusion coefficients and vibrational spectra confirm that NH$_{3}$ molecules dissociate 
between 6000 K and 8000 K (see Supplementary Material). This behaviour reveals that along the Hugoniot the chemistry in ammonia is richer in small molecular species compared to water, which dissociates 
into simple ionic and atomic species at similar shock pressures~\cite{French2010}. 

\begin{figure}[b]
\includegraphics[width=\linewidth]{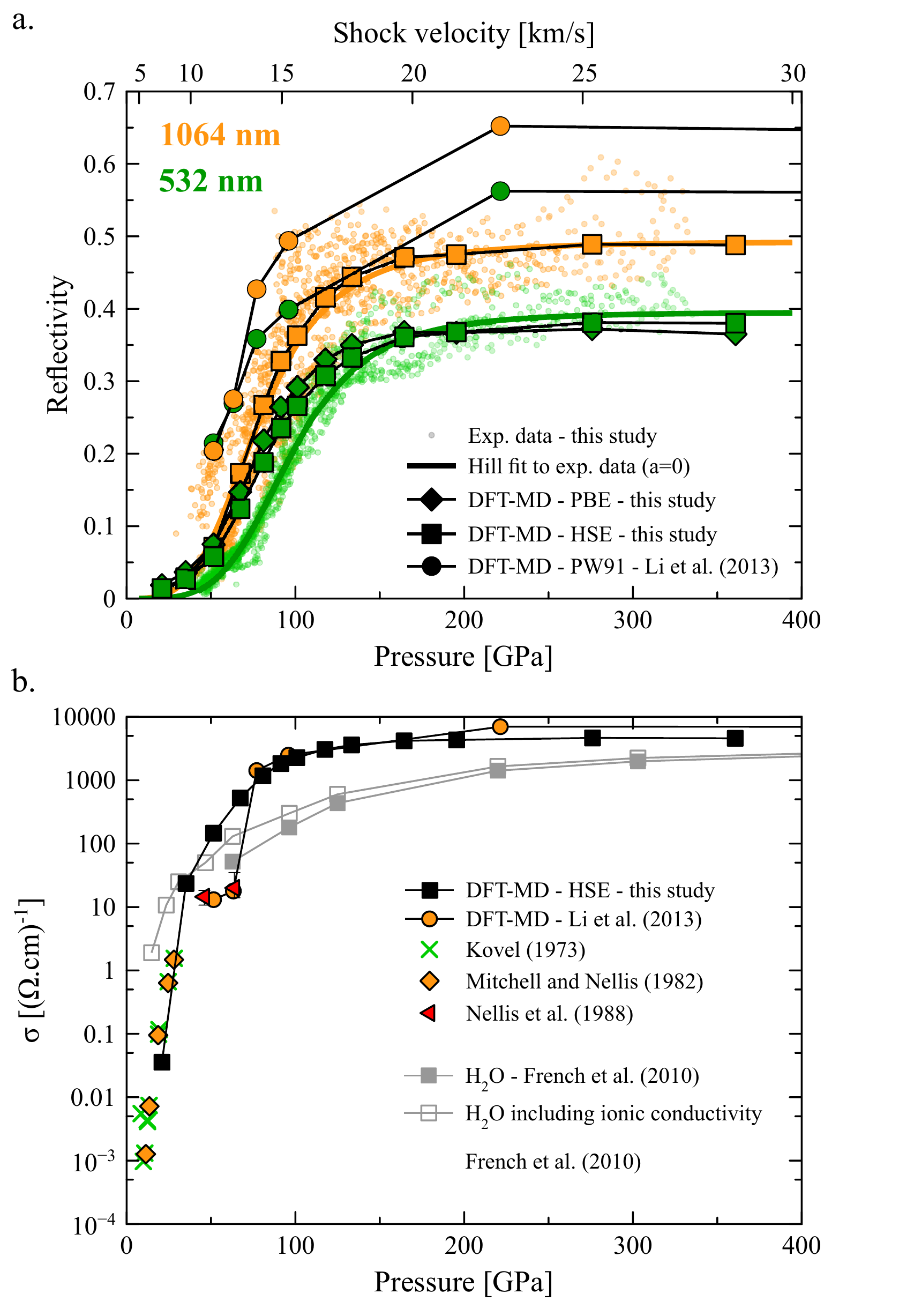}
\caption{
a. Reflectivity at 532~nm (green) and 1064~nm (orange) versus pressure. Small orange and green dots correspond to time-resolved experimental reflectivities measured in 11 decaying shots. Thick orange 
and green lines are Hill fits to the experimental data with the reflectivity imposed to be zero at $U_{s}=0$~km/s. Diamonds and square correspond to the DFT reflectivities computed in this study 
respectively with the PBE and HSE functionals. Large dots show DFT reflectivities from Ref.~\cite{Li2013}. b. Calculated DC electrical conductivity of NH$_{3}$ along the Hugoniot (black squares) compared 
to previous DFT-MD results~\cite{Li2013} and experimental data~\cite{Kovel1973,Mitchell1982,Nellis1997}. The electronic and total electrical conductivity along the principal H$_2$O Hugoniot are also shown 
as gray filled and empty squares, respectively~\cite{French2010}.}
\label{fig: reflectivity}
\end{figure}

Fig.~\ref{fig: reflectivity}a shows the measured reflectivities at 532~nm and 1064~nm, together with the DFT-MD calculated values, using either the PBE~\cite{Perdew1996} or HSE~\cite{Heyd2003,Heyd2006} 
exchange-correlation functionals. Included are also predictions from Ref.~\cite{Li2013}. Our data suggest a gradual rise in shock reflectivity with pressure, with a smooth transition from a low- to a 
highly-reflecting state between $\sim$ 50~GPa and $\sim$ 120~GPa. 

Present calculations are in good agreement with the experimental data. In particular, they match the saturation values at both wavelengths remarkably well. We find a relatively small influence of the 
exchange-correlation functional, due to partial compensation of errors that occurs in the Fresnel formula between the indexes of refraction for the initial ($n_0$) and final ($n$) Hugoniot states (see the 
Supplementary Material). Note that the HSE calculations reproduce the initial index of refraction ($n_{0}^{HSE}=1.34$) much better than the PBE calculations ($n_{0}^{PBE}=1.42$) if compared to the 
experimental value ($n_{0}^{exp}=1.32$). The disagreement between our DFT-MD results and previous calculations is mainly explained by an incorrect setting of $n_0=1$ by Li {\it et al.}~\cite{Li2013}, 
which leads to a shifted onset of reflectivity increase and significantly too high saturation values in their data.
 
Reflectivity is closely related to DC electrical conductivity $\sigma(\omega=0)$, which is a crucial quantity for, e.g., plasma diagnostics and planetary dynamo models. Our {\it ab initio} calculations 
show that the frequency behavior of $\sigma(\omega)$ is very different from a Drude behavior (see Supplemental Material), which is often assumed to derive DC conductivities from optical 
reflectivities~\cite{Celliers2004,Knudson2012,Kimura2015,Millot2018}. The calculated DC electronic conductivity is regarded as highly reliable considering the excellent agreement between {\it ab initio} 
and experimental reflectivity data. It rises systematically with Hugoniot pressure (Fig.~\ref{fig: reflectivity}b). The lowest DC conductivity point at $\sim$ 21~GPa and 2000~K is in good agreement with 
previous experiments~\cite{Kovel1973,Mitchell1982}. Interestingly, we find significantly higher values than those measured in Ref.~\cite{Nellis1997} between 50 GPa and 75 GPa and calculated in 
Li {\it et al.}~\cite{Li2013}. A possible explanation for this deviation may be the higher initial density of $\varrho=0.69$ g/cm$^{3}$ compared to our Hugoniot, so that results from 
Refs. \cite{Li2013,Nellis1997} between 50 GPa and 75 GPa might describe a different phase, most likely the superionic phase~\cite{Bethkenhagen2013}. Above 75 GPa, the values calculated by 
Li {\it et al.}~\cite{Li2013} agree reasonably well with our results. We also note that overall the NH$_3$ conductivity is larger than the one of water along the Hugoniot. Even when the ionic contribution 
is added for water~\cite{French2010}, the ammonia conductivity remains one order of magnitude higher in the 100 GPa regime, due to the different dissociation mechanisms, as discussed above. 


In conclusion, we report novel equation of state measurements of warm dense ammonia in a previously unachieved regime, together with the first experimental evidence of ammonia metallization we observed in 
the reflectivity. These high quality data on such an experimentally challenging compound provide a unique benchmark for our ab initio calculations. The simulations indicate a complex dissociation mechanism 
in ammonia, comprising three regimes: a molecular fluid, where NH$_3$ molecules still persist, a complex fluid above 20~GPa, in which NH$_3$ molecules start to dissociate and N$_2$ and N$_3$ bounded 
molecules are formed, and a completely dissociated plasma state above 90~GPa. The electrical conductivity is surprisingly high and one order of magnitude higher than that of water at $\sim$100~GPa. This 
finding is particularly useful to revisit dynamo models of Uranus and Neptune magnetic fields, as the only existing conductivity data at lower pressures suggest the opposite trend \cite{Nellis1988}. Models 
that entirely disregarded ammonia in the past should now consider the possible contribution of the predicted ammonia-rich layers \cite{NadenRobinson2017} to have better insights of the dynamo process in 
ice giant planets. Future work should extend our study to the superionic domain to estimate the extent and the physical properties of the potential superionic layers within ice giant planets.\\

The authors acknowledge the crucial contribution of the LULI2000 laser and support teams to the success of the experiments. The authors also would like to thank D. Tondelier (LPICM), C. Kaimakian and 
C. Gosmini for technical support, as well as N. Ozaki for useful discussions. This research was supported by the French National Research Agency (ANR) through the projects POMPEI 
(grant no. ANR-16-CE31-0008) and SUPER-ICES (grant ANR-15-CE30-008-01). M.B., M.F., and R.R. gratefully acknowledge support by the DFG within the Research Unit FOR 2440. M.B. was additionally 
supported by the European Union within the Marie Sk{\l}odowska-Curie actions (xICE grant 894725). The DFT-MD calculations were performed at the North-German Supercomputing Alliance (HLRN) facilities. 
J.-A.H. acknowledges support from the Research Council of Norway through its Centres of Excellence funding scheme, project No. 223272. 
The Raman measurements used the spectroscopy platform of IMPMC with the assistance of K. Beneut.


\begin{thebibliography}{83}%
\makeatletter
\providecommand \@ifxundefined [1]{%
 \@ifx{#1\undefined}
}%
\providecommand \@ifnum [1]{%
 \ifnum #1\expandafter \@firstoftwo
 \else \expandafter \@secondoftwo
 \fi
}%
\providecommand \@ifx [1]{%
 \ifx #1\expandafter \@firstoftwo
 \else \expandafter \@secondoftwo
 \fi
}%
\providecommand \natexlab [1]{#1}%
\providecommand \enquote  [1]{``#1''}%
\providecommand \bibnamefont  [1]{#1}%
\providecommand \bibfnamefont [1]{#1}%
\providecommand \citenamefont [1]{#1}%
\providecommand \href@noop [0]{\@secondoftwo}%
\providecommand \href [0]{\begingroup \@sanitize@url \@href}%
\providecommand \@href[1]{\@@startlink{#1}\@@href}%
\providecommand \@@href[1]{\endgroup#1\@@endlink}%
\providecommand \@sanitize@url [0]{\catcode `\\12\catcode `\$12\catcode
  `\&12\catcode `\#12\catcode `\^12\catcode `\_12\catcode `\%12\relax}%
\providecommand \@@startlink[1]{}%
\providecommand \@@endlink[0]{}%
\providecommand \url  [0]{\begingroup\@sanitize@url \@url }%
\providecommand \@url [1]{\endgroup\@href {#1}{\urlprefix }}%
\providecommand \urlprefix  [0]{URL }%
\providecommand \Eprint [0]{\href }%
\providecommand \doibase [0]{http://dx.doi.org/}%
\providecommand \selectlanguage [0]{\@gobble}%
\providecommand \bibinfo  [0]{\@secondoftwo}%
\providecommand \bibfield  [0]{\@secondoftwo}%
\providecommand \translation [1]{[#1]}%
\providecommand \BibitemOpen [0]{}%
\providecommand \bibitemStop [0]{}%
\providecommand \bibitemNoStop [0]{.\EOS\space}%
\providecommand \EOS [0]{\spacefactor3000\relax}%
\providecommand \BibitemShut  [1]{\csname bibitem#1\endcsname}%
\let\auto@bib@innerbib\@empty
\bibitem [{\citenamefont {Liu}\ and\ \citenamefont
  {Tuckerman}(2001)}]{Liu2001}%
  \BibitemOpen
  \bibfield  {author} {\bibinfo {author} {\bibfnamefont {Y.}~\bibnamefont
  {Liu}}\ and\ \bibinfo {author} {\bibfnamefont {M.~E.}\ \bibnamefont
  {Tuckerman}},\ }\href@noop {} {\bibfield  {journal} {\bibinfo  {journal} {J.
  Phys. Chem. B}\ }\textbf {\bibinfo {volume} {105}},\ \bibinfo {pages} {6598}
  (\bibinfo {year} {2001})}\BibitemShut {NoStop}%
\bibitem [{\citenamefont {Fortes}\ \emph {et~al.}(2003)\citenamefont {Fortes},
  \citenamefont {Brodholt}, \citenamefont {Wood},\ and\ \citenamefont
  {Voadlo}}]{Fortes2003}%
  \BibitemOpen
  \bibfield  {author} {\bibinfo {author} {\bibfnamefont {A.~D.}\ \bibnamefont
  {Fortes}}, \bibinfo {author} {\bibfnamefont {J.~P.}\ \bibnamefont
  {Brodholt}}, \bibinfo {author} {\bibfnamefont {I.~G.}\ \bibnamefont {Wood}},
  \ and\ \bibinfo {author} {\bibfnamefont {L.}~\bibnamefont {Voadlo}},\
  }\href@noop {} {\bibfield  {journal} {\bibinfo  {journal} {J. Chem. Phys.}\
  }\textbf {\bibinfo {volume} {118}},\ \bibinfo {pages} {5987} (\bibinfo {year}
  {2003})}\BibitemShut {NoStop}%
\bibitem [{\citenamefont {Hermann}(2017)}]{Hermann2017}%
  \BibitemOpen
  \bibfield  {author} {\bibinfo {author} {\bibfnamefont {A.}~\bibnamefont
  {Hermann}},\ }\href@noop {} {\bibfield  {journal} {\bibinfo  {journal}
  {Reviews in Computational Chemistry}\ }\textbf {\bibinfo {volume} {40}},\
  \bibinfo {pages} {1} (\bibinfo {year} {2017})}\BibitemShut {NoStop}%
\bibitem [{\citenamefont {Cavazzoni}\ \emph {et~al.}(1999)\citenamefont
  {Cavazzoni}, \citenamefont {Chiarotti}, \citenamefont {Scandolo},
  \citenamefont {Tosatti}, \citenamefont {Bernasconi},\ and\ \citenamefont
  {Parrinello}}]{Cavazzoni1999}%
  \BibitemOpen
  \bibfield  {author} {\bibinfo {author} {\bibfnamefont {C.}~\bibnamefont
  {Cavazzoni}}, \bibinfo {author} {\bibfnamefont {G.~L.}\ \bibnamefont
  {Chiarotti}}, \bibinfo {author} {\bibfnamefont {S.}~\bibnamefont {Scandolo}},
  \bibinfo {author} {\bibfnamefont {E.}~\bibnamefont {Tosatti}}, \bibinfo
  {author} {\bibfnamefont {M.}~\bibnamefont {Bernasconi}}, \ and\ \bibinfo
  {author} {\bibfnamefont {M.}~\bibnamefont {Parrinello}},\ }\href {\doibase
  {10.1126/science.283.5398.44}} {\bibfield  {journal} {\bibinfo  {journal}
  {{Science}}\ }\textbf {\bibinfo {volume} {{283}}},\ \bibinfo {pages} {{44}}
  (\bibinfo {year} {{1999}})}\BibitemShut {NoStop}%
\bibitem [{\citenamefont {Pickard}\ and\ \citenamefont
  {Needs}(2008)}]{Pickard2008}%
  \BibitemOpen
  \bibfield  {author} {\bibinfo {author} {\bibfnamefont {C.~J.}\ \bibnamefont
  {Pickard}}\ and\ \bibinfo {author} {\bibfnamefont {R.~J.}\ \bibnamefont
  {Needs}},\ }\href {\doibase {10.1038/nmat2261}} {\bibfield  {journal}
  {\bibinfo  {journal} {{Nature Materials}}\ }\textbf {\bibinfo {volume}
  {{7}}},\ \bibinfo {pages} {{775}} (\bibinfo {year} {{2008}})}\BibitemShut
  {NoStop}%
\bibitem [{\citenamefont {{Bethkenhagen}}\ \emph {et~al.}(2013)\citenamefont
  {{Bethkenhagen}}, \citenamefont {{French}},\ and\ \citenamefont
  {{Redmer}}}]{Bethkenhagen2013}%
  \BibitemOpen
  \bibfield  {author} {\bibinfo {author} {\bibfnamefont {M.}~\bibnamefont
  {{Bethkenhagen}}}, \bibinfo {author} {\bibfnamefont {M.}~\bibnamefont
  {{French}}}, \ and\ \bibinfo {author} {\bibfnamefont {R.}~\bibnamefont
  {{Redmer}}},\ }\href@noop {} {\bibfield  {journal} {\bibinfo  {journal}
  {Journal of Chemical Physics}\ }\textbf {\bibinfo {volume} {138}},\ \bibinfo
  {pages} {234504} (\bibinfo {year} {2013})}\BibitemShut {NoStop}%
\bibitem [{\citenamefont {Hubbard}\ and\ \citenamefont
  {MacFarlane}(1980)}]{Hubbard1980}%
  \BibitemOpen
  \bibfield  {author} {\bibinfo {author} {\bibfnamefont {W.~B.}\ \bibnamefont
  {Hubbard}}\ and\ \bibinfo {author} {\bibfnamefont {J.~J.}\ \bibnamefont
  {MacFarlane}},\ }\href@noop {} {\bibfield  {journal} {\bibinfo  {journal} {J.
  Geophys. Res.}\ }\textbf {\bibinfo {volume} {85}},\ \bibinfo {pages} {225}
  (\bibinfo {year} {1980})}\BibitemShut {NoStop}%
\bibitem [{\citenamefont {Hubbard}\ \emph {et~al.}(1991)\citenamefont
  {Hubbard}, \citenamefont {Nellis}, \citenamefont {Mitchell}, \citenamefont
  {Holmes}, \citenamefont {Limaye},\ and\ \citenamefont
  {McCandless}}]{Hubbard1991}%
  \BibitemOpen
  \bibfield  {author} {\bibinfo {author} {\bibfnamefont {W.~B.}\ \bibnamefont
  {Hubbard}}, \bibinfo {author} {\bibfnamefont {W.~J.}\ \bibnamefont {Nellis}},
  \bibinfo {author} {\bibfnamefont {A.~C.}\ \bibnamefont {Mitchell}}, \bibinfo
  {author} {\bibfnamefont {N.~C.}\ \bibnamefont {Holmes}}, \bibinfo {author}
  {\bibfnamefont {S.~S.}\ \bibnamefont {Limaye}}, \ and\ \bibinfo {author}
  {\bibfnamefont {P.~C.}\ \bibnamefont {McCandless}},\ }\href {\doibase
  {10.1126/science.253.5020.648}} {\bibfield  {journal} {\bibinfo  {journal}
  {{Science}}\ }\textbf {\bibinfo {volume} {{253}}},\ \bibinfo {pages} {{648}}
  (\bibinfo {year} {{1991}})}\BibitemShut {NoStop}%
\bibitem [{\citenamefont {Journaux}\ \emph {et~al.}(2020)\citenamefont
  {Journaux}, \citenamefont {Kalousova}, \citenamefont {Sotin}, \citenamefont
  {Tobie}, \citenamefont {Vance}, \citenamefont {Saur}, \citenamefont
  {Bollengier}, \citenamefont {Noack}, \citenamefont {Ruckriemen-Bez},
  \citenamefont {Van~Hoolst}, \citenamefont {Soderlund},\ and\ \citenamefont
  {Brown}}]{Journaux2020}%
  \BibitemOpen
  \bibfield  {author} {\bibinfo {author} {\bibfnamefont {B.}~\bibnamefont
  {Journaux}}, \bibinfo {author} {\bibfnamefont {K.}~\bibnamefont {Kalousova}},
  \bibinfo {author} {\bibfnamefont {C.}~\bibnamefont {Sotin}}, \bibinfo
  {author} {\bibfnamefont {G.}~\bibnamefont {Tobie}}, \bibinfo {author}
  {\bibfnamefont {S.}~\bibnamefont {Vance}}, \bibinfo {author} {\bibfnamefont
  {J.}~\bibnamefont {Saur}}, \bibinfo {author} {\bibfnamefont {O.}~\bibnamefont
  {Bollengier}}, \bibinfo {author} {\bibfnamefont {L.}~\bibnamefont {Noack}},
  \bibinfo {author} {\bibfnamefont {T.}~\bibnamefont {Ruckriemen-Bez}},
  \bibinfo {author} {\bibfnamefont {T.}~\bibnamefont {Van~Hoolst}}, \bibinfo
  {author} {\bibfnamefont {K.~M.}\ \bibnamefont {Soderlund}}, \ and\ \bibinfo
  {author} {\bibfnamefont {J.~M.}\ \bibnamefont {Brown}},\ }\href@noop {}
  {\bibfield  {journal} {\bibinfo  {journal} {Space Science Review}\ }\textbf
  {\bibinfo {volume} {216}},\ \bibinfo {pages} {7} (\bibinfo {year}
  {2020})}\BibitemShut {NoStop}%
\bibitem [{\citenamefont {Helled}\ \emph {et~al.}(2020)\citenamefont {Helled},
  \citenamefont {Nettelmann},\ and\ \citenamefont {Guillot}}]{Helled2020}%
  \BibitemOpen
  \bibfield  {author} {\bibinfo {author} {\bibfnamefont {R.}~\bibnamefont
  {Helled}}, \bibinfo {author} {\bibfnamefont {N.}~\bibnamefont {Nettelmann}},
  \ and\ \bibinfo {author} {\bibfnamefont {T.}~\bibnamefont {Guillot}},\
  }\href@noop {} {\bibfield  {journal} {\bibinfo  {journal} {Space Sci. Rev.}\
  }\textbf {\bibinfo {volume} {216}},\ \bibinfo {pages} {38} (\bibinfo {year}
  {2020})}\BibitemShut {NoStop}%
\bibitem [{\citenamefont {Borucki}\ and\ \citenamefont
  {et~al.}(2011)}]{Borucki2011}%
  \BibitemOpen
  \bibfield  {author} {\bibinfo {author} {\bibfnamefont {W.~J.}\ \bibnamefont
  {Borucki}}\ and\ \bibinfo {author} {\bibnamefont {et~al.}},\ }\href@noop {}
  {\bibfield  {journal} {\bibinfo  {journal} {Astrophysical Journal}\ }\textbf
  {\bibinfo {volume} {736}},\ \bibinfo {pages} {19} (\bibinfo {year}
  {2011})}\BibitemShut {NoStop}%
\bibitem [{\citenamefont {{Fressin}}\ \emph {et~al.}(2013)\citenamefont
  {{Fressin}}, \citenamefont {{Torres}}, \citenamefont {{Charbonneau}},
  \citenamefont {{Bryson}}, \citenamefont {{Christiansen}}, \citenamefont
  {{Dressing}}, \citenamefont {{Jenkins}}, \citenamefont {{Walkowicz}},\ and\
  \citenamefont {{Batalha}}}]{Fressin2013}%
  \BibitemOpen
  \bibfield  {author} {\bibinfo {author} {\bibfnamefont {F.}~\bibnamefont
  {{Fressin}}}, \bibinfo {author} {\bibfnamefont {G.}~\bibnamefont {{Torres}}},
  \bibinfo {author} {\bibfnamefont {D.}~\bibnamefont {{Charbonneau}}}, \bibinfo
  {author} {\bibfnamefont {S.~T.}\ \bibnamefont {{Bryson}}}, \bibinfo {author}
  {\bibfnamefont {J.}~\bibnamefont {{Christiansen}}}, \bibinfo {author}
  {\bibfnamefont {C.~D.}\ \bibnamefont {{Dressing}}}, \bibinfo {author}
  {\bibfnamefont {J.~M.}\ \bibnamefont {{Jenkins}}}, \bibinfo {author}
  {\bibfnamefont {L.~M.}\ \bibnamefont {{Walkowicz}}}, \ and\ \bibinfo {author}
  {\bibfnamefont {N.~M.}\ \bibnamefont {{Batalha}}},\ }\href@noop {} {\bibfield
   {journal} {\bibinfo  {journal} {Astrophysical Journal}\ }\textbf {\bibinfo
  {volume} {766}},\ \bibinfo {pages} {81} (\bibinfo {year} {2013})}\BibitemShut
  {NoStop}%
\bibitem [{\citenamefont {Zeng}\ \emph {et~al.}(2019)\citenamefont {Zeng},
  \citenamefont {Jacobsen}, \citenamefont {Sasselov}, \citenamefont {Petaev},
  \citenamefont {Vanderburg}, \citenamefont {Lopez-Morales}, \citenamefont
  {Perez-Mercader}, \citenamefont {Mattsson}, \citenamefont {Li}, \citenamefont
  {Heising}, \citenamefont {Bonomo}, \citenamefont {Damasso}, \citenamefont
  {Berger}, \citenamefont {Cao}, \citenamefont {Levi},\ and\ \citenamefont
  {Wordsworth}}]{Zeng2019}%
  \BibitemOpen
  \bibfield  {author} {\bibinfo {author} {\bibfnamefont {L.}~\bibnamefont
  {Zeng}}, \bibinfo {author} {\bibfnamefont {S.~B.}\ \bibnamefont {Jacobsen}},
  \bibinfo {author} {\bibfnamefont {D.~D.}\ \bibnamefont {Sasselov}}, \bibinfo
  {author} {\bibfnamefont {M.~I.}\ \bibnamefont {Petaev}}, \bibinfo {author}
  {\bibfnamefont {A.}~\bibnamefont {Vanderburg}}, \bibinfo {author}
  {\bibfnamefont {M.}~\bibnamefont {Lopez-Morales}}, \bibinfo {author}
  {\bibfnamefont {J.}~\bibnamefont {Perez-Mercader}}, \bibinfo {author}
  {\bibfnamefont {T.~R.}\ \bibnamefont {Mattsson}}, \bibinfo {author}
  {\bibfnamefont {G.}~\bibnamefont {Li}}, \bibinfo {author} {\bibfnamefont
  {M.~Z.}\ \bibnamefont {Heising}}, \bibinfo {author} {\bibfnamefont {A.~S.}\
  \bibnamefont {Bonomo}}, \bibinfo {author} {\bibfnamefont {M.}~\bibnamefont
  {Damasso}}, \bibinfo {author} {\bibfnamefont {T.~A.}\ \bibnamefont {Berger}},
  \bibinfo {author} {\bibfnamefont {H.}~\bibnamefont {Cao}}, \bibinfo {author}
  {\bibfnamefont {A.}~\bibnamefont {Levi}}, \ and\ \bibinfo {author}
  {\bibfnamefont {R.~D.}\ \bibnamefont {Wordsworth}},\ }\href {\doibase
  10.1073/pnas.1812905116} {\bibfield  {journal} {\bibinfo  {journal}
  {Proceedings of the National Academy of Sciences}\ }\textbf {\bibinfo
  {volume} {116}},\ \bibinfo {pages} {9723} (\bibinfo {year}
  {2019})}\BibitemShut {NoStop}%
\bibitem [{\citenamefont {Redmer}\ \emph {et~al.}(2011)\citenamefont {Redmer},
  \citenamefont {Mattsson}, \citenamefont {Nettelmann},\ and\ \citenamefont
  {French}}]{Redmer2011}%
  \BibitemOpen
  \bibfield  {author} {\bibinfo {author} {\bibfnamefont {R.}~\bibnamefont
  {Redmer}}, \bibinfo {author} {\bibfnamefont {T.~R.}\ \bibnamefont
  {Mattsson}}, \bibinfo {author} {\bibfnamefont {N.}~\bibnamefont
  {Nettelmann}}, \ and\ \bibinfo {author} {\bibfnamefont {M.}~\bibnamefont
  {French}},\ }\href {\doibase {10.1016/j.icarus.2010.08.008}} {\bibfield
  {journal} {\bibinfo  {journal} {{Icarus}}\ }\textbf {\bibinfo {volume}
  {{211}}},\ \bibinfo {pages} {{798}} (\bibinfo {year} {{2011}})}\BibitemShut
  {NoStop}%
\bibitem [{\citenamefont {Helled}\ \emph {et~al.}(2011)\citenamefont {Helled},
  \citenamefont {Anderson}, \citenamefont {Podolak},\ and\ \citenamefont
  {Schubert}}]{Helled2011}%
  \BibitemOpen
  \bibfield  {author} {\bibinfo {author} {\bibfnamefont {R.}~\bibnamefont
  {Helled}}, \bibinfo {author} {\bibfnamefont {J.}~\bibnamefont {Anderson}},
  \bibinfo {author} {\bibfnamefont {M.}~\bibnamefont {Podolak}}, \ and\
  \bibinfo {author} {\bibfnamefont {G.}~\bibnamefont {Schubert}},\ }\href@noop
  {} {\bibfield  {journal} {\bibinfo  {journal} {ApJ}\ }\textbf {\bibinfo
  {volume} {726}},\ \bibinfo {pages} {A15} (\bibinfo {year}
  {2011})}\BibitemShut {NoStop}%
\bibitem [{\citenamefont {Nettelmann}\ \emph {et~al.}(2013)\citenamefont
  {Nettelmann}, \citenamefont {Helled}, \citenamefont {Fortney},\ and\
  \citenamefont {Redmer}}]{Nettelmann2013}%
  \BibitemOpen
  \bibfield  {author} {\bibinfo {author} {\bibfnamefont {N.}~\bibnamefont
  {Nettelmann}}, \bibinfo {author} {\bibfnamefont {R.}~\bibnamefont {Helled}},
  \bibinfo {author} {\bibfnamefont {J.}~\bibnamefont {Fortney}}, \ and\
  \bibinfo {author} {\bibfnamefont {R.}~\bibnamefont {Redmer}},\ }\href@noop {}
  {\bibfield  {journal} {\bibinfo  {journal} {Planetary and Space Science}\
  }\textbf {\bibinfo {volume} {77}},\ \bibinfo {pages} {143} (\bibinfo {year}
  {2013})}\BibitemShut {NoStop}%
\bibitem [{\citenamefont {Bethkenhagen}\ \emph {et~al.}(2017)\citenamefont
  {Bethkenhagen}, \citenamefont {Meyer}, \citenamefont {Hamel}, \citenamefont
  {Nettelmann}, \citenamefont {French}, \citenamefont {Scheibe}, \citenamefont
  {Ticknor}, \citenamefont {Collins}, \citenamefont {Kress}, \citenamefont
  {Fortney},\ and\ \citenamefont {Redmer}}]{Bethkenhagen2017}%
  \BibitemOpen
  \bibfield  {author} {\bibinfo {author} {\bibfnamefont {M.}~\bibnamefont
  {Bethkenhagen}}, \bibinfo {author} {\bibfnamefont {E.~R.}\ \bibnamefont
  {Meyer}}, \bibinfo {author} {\bibfnamefont {S.}~\bibnamefont {Hamel}},
  \bibinfo {author} {\bibfnamefont {N.}~\bibnamefont {Nettelmann}}, \bibinfo
  {author} {\bibfnamefont {M.}~\bibnamefont {French}}, \bibinfo {author}
  {\bibfnamefont {L.}~\bibnamefont {Scheibe}}, \bibinfo {author} {\bibfnamefont
  {C.}~\bibnamefont {Ticknor}}, \bibinfo {author} {\bibfnamefont {L.~A.}\
  \bibnamefont {Collins}}, \bibinfo {author} {\bibfnamefont {J.~D.}\
  \bibnamefont {Kress}}, \bibinfo {author} {\bibfnamefont {J.~J.}\ \bibnamefont
  {Fortney}}, \ and\ \bibinfo {author} {\bibfnamefont {R.}~\bibnamefont
  {Redmer}},\ }\href@noop {} {\bibfield  {journal} {\bibinfo  {journal} {The
  Astrophysical Journal}\ }\textbf {\bibinfo {volume} {848}},\ \bibinfo {pages}
  {67} (\bibinfo {year} {2017})}\BibitemShut {NoStop}%
\bibitem [{\citenamefont {Scheibe}\ \emph {et~al.}(2019)\citenamefont
  {Scheibe}, \citenamefont {Nettelmann},\ and\ \citenamefont
  {Redmer}}]{Scheibe2019}%
  \BibitemOpen
  \bibfield  {author} {\bibinfo {author} {\bibfnamefont {L.}~\bibnamefont
  {Scheibe}}, \bibinfo {author} {\bibfnamefont {N.}~\bibnamefont {Nettelmann}},
  \ and\ \bibinfo {author} {\bibfnamefont {R.}~\bibnamefont {Redmer}},\
  }\href@noop {} {\bibfield  {journal} {\bibinfo  {journal} {Astronomy \&
  Astrophysics}\ }\textbf {\bibinfo {volume} {632}},\ \bibinfo {pages} {A70}
  (\bibinfo {year} {2019})}\BibitemShut {NoStop}%
\bibitem [{\citenamefont {Nettelmann}\ \emph {et~al.}(2016)\citenamefont
  {Nettelmann}, \citenamefont {Wang}, \citenamefont {Fortney}, \citenamefont
  {Hamel}, \citenamefont {Yellamilli}, \citenamefont {Bethkenhagen},\ and\
  \citenamefont {Redmer}}]{Nettelmann2016}%
  \BibitemOpen
  \bibfield  {author} {\bibinfo {author} {\bibfnamefont {N.}~\bibnamefont
  {Nettelmann}}, \bibinfo {author} {\bibfnamefont {K.}~\bibnamefont {Wang}},
  \bibinfo {author} {\bibfnamefont {J.}~\bibnamefont {Fortney}}, \bibinfo
  {author} {\bibfnamefont {S.}~\bibnamefont {Hamel}}, \bibinfo {author}
  {\bibfnamefont {S.}~\bibnamefont {Yellamilli}}, \bibinfo {author}
  {\bibfnamefont {M.}~\bibnamefont {Bethkenhagen}}, \ and\ \bibinfo {author}
  {\bibfnamefont {R.}~\bibnamefont {Redmer}},\ }\href@noop {} {\bibfield
  {journal} {\bibinfo  {journal} {Icarus}\ }\textbf {\bibinfo {volume} {275}},\
  \bibinfo {pages} {107} (\bibinfo {year} {2016})}\BibitemShut {NoStop}%
\bibitem [{\citenamefont {Podolak}\ \emph {et~al.}(2019)\citenamefont
  {Podolak}, \citenamefont {Helled},\ and\ \citenamefont
  {Schubert}}]{Podolak2019}%
  \BibitemOpen
  \bibfield  {author} {\bibinfo {author} {\bibfnamefont {M.}~\bibnamefont
  {Podolak}}, \bibinfo {author} {\bibfnamefont {R.}~\bibnamefont {Helled}}, \
  and\ \bibinfo {author} {\bibfnamefont {G.}~\bibnamefont {Schubert}},\
  }\href@noop {} {\bibfield  {journal} {\bibinfo  {journal} {Monthly Notices of
  the Royal Astronomical Society}\ }\textbf {\bibinfo {volume} {487}},\
  \bibinfo {pages} {2653} (\bibinfo {year} {2019})}\BibitemShut {NoStop}%
\bibitem [{\citenamefont {Vazan}\ and\ \citenamefont
  {Helled}(2020)}]{Vazan2020}%
  \BibitemOpen
  \bibfield  {author} {\bibinfo {author} {\bibfnamefont {A.}~\bibnamefont
  {Vazan}}\ and\ \bibinfo {author} {\bibfnamefont {R.}~\bibnamefont {Helled}},\
  }\href@noop {} {\bibfield  {journal} {\bibinfo  {journal} {Astronomy \&
  Astrophysics}\ }\textbf {\bibinfo {volume} {633}},\ \bibinfo {pages} {A50}
  (\bibinfo {year} {2020})}\BibitemShut {NoStop}%
\bibitem [{\citenamefont {Li}\ \emph {et~al.}(2013)\citenamefont {Li},
  \citenamefont {Zhang},\ and\ \citenamefont {Yan}}]{Li2013}%
  \BibitemOpen
  \bibfield  {author} {\bibinfo {author} {\bibfnamefont {D.~F.}\ \bibnamefont
  {Li}}, \bibinfo {author} {\bibfnamefont {P.}~\bibnamefont {Zhang}}, \ and\
  \bibinfo {author} {\bibfnamefont {J.}~\bibnamefont {Yan}},\ }\href@noop {}
  {\bibfield  {journal} {\bibinfo  {journal} {Journal of Chemical Physics}\
  }\textbf {\bibinfo {volume} {139}},\ \bibinfo {pages} {134505} (\bibinfo
  {year} {2013})}\BibitemShut {NoStop}%
\bibitem [{\citenamefont {Li}\ \emph {et~al.}(2017)\citenamefont {Li},
  \citenamefont {Wang}, \citenamefont {Yan}, \citenamefont {Fu},\ and\
  \citenamefont {Zhang}}]{Li2017}%
  \BibitemOpen
  \bibfield  {author} {\bibinfo {author} {\bibfnamefont {D.~F.}\ \bibnamefont
  {Li}}, \bibinfo {author} {\bibfnamefont {C.}~\bibnamefont {Wang}}, \bibinfo
  {author} {\bibfnamefont {J.}~\bibnamefont {Yan}}, \bibinfo {author}
  {\bibfnamefont {Z.~G.}\ \bibnamefont {Fu}}, \ and\ \bibinfo {author}
  {\bibfnamefont {P.}~\bibnamefont {Zhang}},\ }\href@noop {} {\bibfield
  {journal} {\bibinfo  {journal} {Scientific Reports}\ }\textbf {\bibinfo
  {volume} {7}},\ \bibinfo {pages} {12338} (\bibinfo {year}
  {2017})}\BibitemShut {NoStop}%
\bibitem [{\citenamefont {Nettelmann}\ \emph {et~al.}(2010)\citenamefont
  {Nettelmann}, \citenamefont {Kramm}, \citenamefont {Redmer},\ and\
  \citenamefont {Neuh\"auser}}]{Nettelmann2010}%
  \BibitemOpen
  \bibfield  {author} {\bibinfo {author} {\bibfnamefont {N.}~\bibnamefont
  {Nettelmann}}, \bibinfo {author} {\bibfnamefont {U.}~\bibnamefont {Kramm}},
  \bibinfo {author} {\bibfnamefont {R.}~\bibnamefont {Redmer}}, \ and\ \bibinfo
  {author} {\bibnamefont {Neuh\"auser}},\ }\href@noop {} {\bibfield  {journal}
  {\bibinfo  {journal} {Astronomy \& Astrophysics}\ }\textbf {\bibinfo {volume}
  {523}} (\bibinfo {year} {2010})}\BibitemShut {NoStop}%
\bibitem [{\citenamefont {Guarguaglini}\ \emph {et~al.}(2019)\citenamefont
  {Guarguaglini}, \citenamefont {Hernandez}, \citenamefont {Okuchi},
  \citenamefont {Barroso}, \citenamefont {Benuzzi-Mounaix}, \citenamefont
  {Bethkenhagen}, \citenamefont {Bolis}, \citenamefont {Brambrink},
  \citenamefont {French}, \citenamefont {Fujimoto}, \citenamefont {Kodama},
  \citenamefont {Koenig}, \citenamefont {Lefevre}, \citenamefont {Miyanishi},
  \citenamefont {Ozaki}, \citenamefont {Redmer}, \citenamefont {Sano},
  \citenamefont {Umeda}, \citenamefont {Vinci},\ and\ \citenamefont
  {Ravasio}}]{Guarguaglini2019}%
  \BibitemOpen
  \bibfield  {author} {\bibinfo {author} {\bibfnamefont {M.}~\bibnamefont
  {Guarguaglini}}, \bibinfo {author} {\bibfnamefont {J.~A.}\ \bibnamefont
  {Hernandez}}, \bibinfo {author} {\bibfnamefont {T.}~\bibnamefont {Okuchi}},
  \bibinfo {author} {\bibfnamefont {P.}~\bibnamefont {Barroso}}, \bibinfo
  {author} {\bibfnamefont {A.}~\bibnamefont {Benuzzi-Mounaix}}, \bibinfo
  {author} {\bibfnamefont {M.}~\bibnamefont {Bethkenhagen}}, \bibinfo {author}
  {\bibfnamefont {R.}~\bibnamefont {Bolis}}, \bibinfo {author} {\bibfnamefont
  {E.}~\bibnamefont {Brambrink}}, \bibinfo {author} {\bibfnamefont
  {M.}~\bibnamefont {French}}, \bibinfo {author} {\bibfnamefont
  {Y.}~\bibnamefont {Fujimoto}}, \bibinfo {author} {\bibfnamefont
  {R.}~\bibnamefont {Kodama}}, \bibinfo {author} {\bibfnamefont
  {M.}~\bibnamefont {Koenig}}, \bibinfo {author} {\bibfnamefont
  {F.}~\bibnamefont {Lefevre}}, \bibinfo {author} {\bibfnamefont
  {K.}~\bibnamefont {Miyanishi}}, \bibinfo {author} {\bibfnamefont
  {N.}~\bibnamefont {Ozaki}}, \bibinfo {author} {\bibfnamefont
  {R.}~\bibnamefont {Redmer}}, \bibinfo {author} {\bibfnamefont
  {T.}~\bibnamefont {Sano}}, \bibinfo {author} {\bibfnamefont {Y.}~\bibnamefont
  {Umeda}}, \bibinfo {author} {\bibfnamefont {T.}~\bibnamefont {Vinci}}, \ and\
  \bibinfo {author} {\bibfnamefont {A.}~\bibnamefont {Ravasio}},\ }\href@noop
  {} {\bibfield  {journal} {\bibinfo  {journal} {Scientific Reports}\ }\textbf
  {\bibinfo {volume} {9}},\ \bibinfo {pages} {10155} (\bibinfo {year}
  {2019})}\BibitemShut {NoStop}%
\bibitem [{\citenamefont {Stevenson}(2010)}]{Stevenson2010}%
  \BibitemOpen
  \bibfield  {author} {\bibinfo {author} {\bibfnamefont {D.~J.}\ \bibnamefont
  {Stevenson}},\ }\href {\doibase {10.1007/s11214-009-9572-z}} {\bibfield
  {journal} {\bibinfo  {journal} {{Space Science Reviews}}\ }\textbf {\bibinfo
  {volume} {{152}}},\ \bibinfo {pages} {{651}} (\bibinfo {year}
  {{2010}})}\BibitemShut {NoStop}%
\bibitem [{\citenamefont {Stanley}\ and\ \citenamefont
  {Bloxham}(2006)}]{Stanley2006}%
  \BibitemOpen
  \bibfield  {author} {\bibinfo {author} {\bibfnamefont {S.}~\bibnamefont
  {Stanley}}\ and\ \bibinfo {author} {\bibfnamefont {J.}~\bibnamefont
  {Bloxham}},\ }\href {\doibase {10.1016/j.icarus.2006.05.005}} {\bibfield
  {journal} {\bibinfo  {journal} {{Icarus}}\ }\textbf {\bibinfo {volume}
  {{184}}},\ \bibinfo {pages} {{556}} (\bibinfo {year} {{2006}})}\BibitemShut
  {NoStop}%
\bibitem [{\citenamefont {Soderlund}\ \emph {et~al.}(2013)\citenamefont
  {Soderlund}, \citenamefont {Heimpel}, \citenamefont {King},\ and\
  \citenamefont {Aurnou}}]{Soderlund2013}%
  \BibitemOpen
  \bibfield  {author} {\bibinfo {author} {\bibfnamefont {K.~M.}\ \bibnamefont
  {Soderlund}}, \bibinfo {author} {\bibfnamefont {M.~H.}\ \bibnamefont
  {Heimpel}}, \bibinfo {author} {\bibfnamefont {E.~M.}\ \bibnamefont {King}}, \
  and\ \bibinfo {author} {\bibfnamefont {J.~M.}\ \bibnamefont {Aurnou}},\
  }\href@noop {} {\bibfield  {journal} {\bibinfo  {journal} {Icarus}\ }\textbf
  {\bibinfo {volume} {224}},\ \bibinfo {pages} {97} (\bibinfo {year}
  {2013})}\BibitemShut {NoStop}%
\bibitem [{\citenamefont {Ninet}\ and\ \citenamefont
  {Datchi}(2008)}]{Ninet2008}%
  \BibitemOpen
  \bibfield  {author} {\bibinfo {author} {\bibfnamefont {S.}~\bibnamefont
  {Ninet}}\ and\ \bibinfo {author} {\bibfnamefont {F.}~\bibnamefont {Datchi}},\
  }\href {\doibase {10.1063/1.2903491}} {\bibfield  {journal} {\bibinfo
  {journal} {{The Journal of Chemical Physics}}\ }\textbf {\bibinfo {volume}
  {{128}}},\ \bibinfo {pages} {154508} (\bibinfo {year} {{2008}})}\BibitemShut
  {NoStop}%
\bibitem [{\citenamefont {Li}\ \emph {et~al.}(2009)\citenamefont {Li},
  \citenamefont {Li}, \citenamefont {Cui}, \citenamefont {Cui}, \citenamefont
  {He}, \citenamefont {Zhou},\ and\ \citenamefont {Zou}}]{Li2009}%
  \BibitemOpen
  \bibfield  {author} {\bibinfo {author} {\bibfnamefont {F.}~\bibnamefont
  {Li}}, \bibinfo {author} {\bibfnamefont {M.}~\bibnamefont {Li}}, \bibinfo
  {author} {\bibfnamefont {Q.}~\bibnamefont {Cui}}, \bibinfo {author}
  {\bibfnamefont {T.}~\bibnamefont {Cui}}, \bibinfo {author} {\bibfnamefont
  {Z.}~\bibnamefont {He}}, \bibinfo {author} {\bibfnamefont {Q.}~\bibnamefont
  {Zhou}}, \ and\ \bibinfo {author} {\bibfnamefont {G.}~\bibnamefont {Zou}},\
  }\href {\doibase {10.1063/1.3223549}} {\bibfield  {journal} {\bibinfo
  {journal} {{The Journal of Chemical Physics}}\ }\textbf {\bibinfo {volume}
  {{131}}} (\bibinfo {year} {{2009}}),\ {10.1063/1.3223549}}\BibitemShut
  {NoStop}%
\bibitem [{\citenamefont {Ninet}\ \emph {et~al.}(2012)\citenamefont {Ninet},
  \citenamefont {Datchi},\ and\ \citenamefont {Saitta}}]{Ninet2012}%
  \BibitemOpen
  \bibfield  {author} {\bibinfo {author} {\bibfnamefont {S.}~\bibnamefont
  {Ninet}}, \bibinfo {author} {\bibfnamefont {F.}~\bibnamefont {Datchi}}, \
  and\ \bibinfo {author} {\bibfnamefont {A.~M.}\ \bibnamefont {Saitta}},\
  }\href {\doibase 10.1103/PhysRevLett.108.165702} {\bibfield  {journal}
  {\bibinfo  {journal} {Physical Review Letters}\ }\textbf {\bibinfo {volume}
  {108}},\ \bibinfo {pages} {165702} (\bibinfo {year} {2012})}\BibitemShut
  {NoStop}%
\bibitem [{\citenamefont {Ninet}\ \emph {et~al.}(2014)\citenamefont {Ninet},
  \citenamefont {Datchi}, \citenamefont {Dumas}, \citenamefont {Mezouar},
  \citenamefont {Garbarino}, \citenamefont {Mafety}, \citenamefont {Pickard},
  \citenamefont {Needs},\ and\ \citenamefont {Saitta}}]{Ninet2014}%
  \BibitemOpen
  \bibfield  {author} {\bibinfo {author} {\bibfnamefont {S.}~\bibnamefont
  {Ninet}}, \bibinfo {author} {\bibfnamefont {F.}~\bibnamefont {Datchi}},
  \bibinfo {author} {\bibfnamefont {P.}~\bibnamefont {Dumas}}, \bibinfo
  {author} {\bibfnamefont {M.}~\bibnamefont {Mezouar}}, \bibinfo {author}
  {\bibfnamefont {G.}~\bibnamefont {Garbarino}}, \bibinfo {author}
  {\bibfnamefont {A.}~\bibnamefont {Mafety}}, \bibinfo {author} {\bibfnamefont
  {C.~J.}\ \bibnamefont {Pickard}}, \bibinfo {author} {\bibfnamefont {R.~J.}\
  \bibnamefont {Needs}}, \ and\ \bibinfo {author} {\bibfnamefont {A.~M.}\
  \bibnamefont {Saitta}},\ }\href@noop {} {\bibfield  {journal} {\bibinfo
  {journal} {Physical Review B}\ }\textbf {\bibinfo {volume} {89}},\ \bibinfo
  {pages} {174103} (\bibinfo {year} {2014})}\BibitemShut {NoStop}%
\bibitem [{\citenamefont {Queyroux}\ \emph
  {et~al.}(2019{\natexlab{a}})\citenamefont {Queyroux}, \citenamefont {Ninet},
  \citenamefont {Weck}, \citenamefont {Garbarino}, \citenamefont {Plisson},
  \citenamefont {Mezouar},\ and\ \citenamefont {Datchi}}]{Queyroux2019}%
  \BibitemOpen
  \bibfield  {author} {\bibinfo {author} {\bibfnamefont {J.~A.}\ \bibnamefont
  {Queyroux}}, \bibinfo {author} {\bibfnamefont {S.}~\bibnamefont {Ninet}},
  \bibinfo {author} {\bibfnamefont {G.}~\bibnamefont {Weck}}, \bibinfo {author}
  {\bibfnamefont {G.}~\bibnamefont {Garbarino}}, \bibinfo {author}
  {\bibfnamefont {T.}~\bibnamefont {Plisson}}, \bibinfo {author} {\bibfnamefont
  {M.}~\bibnamefont {Mezouar}}, \ and\ \bibinfo {author} {\bibfnamefont
  {F.}~\bibnamefont {Datchi}},\ }\href@noop {} {\bibfield  {journal} {\bibinfo
  {journal} {Physical Review B}\ }\textbf {\bibinfo {volume} {99}},\ \bibinfo
  {pages} {134107} (\bibinfo {year} {2019}{\natexlab{a}})}\BibitemShut
  {NoStop}%
\bibitem [{\citenamefont {Queyroux}\ \emph
  {et~al.}(2019{\natexlab{b}})\citenamefont {Queyroux}, \citenamefont {Ninet},
  \citenamefont {Weck}, \citenamefont {Garbarino}, \citenamefont {Mezouar},\
  and\ \citenamefont {Datchi}}]{Queyroux2019b}%
  \BibitemOpen
  \bibfield  {author} {\bibinfo {author} {\bibfnamefont {J.~A.}\ \bibnamefont
  {Queyroux}}, \bibinfo {author} {\bibfnamefont {S.}~\bibnamefont {Ninet}},
  \bibinfo {author} {\bibfnamefont {G.}~\bibnamefont {Weck}}, \bibinfo {author}
  {\bibfnamefont {G.}~\bibnamefont {Garbarino}}, \bibinfo {author}
  {\bibfnamefont {M.}~\bibnamefont {Mezouar}}, \ and\ \bibinfo {author}
  {\bibfnamefont {F.}~\bibnamefont {Datchi}},\ }\href@noop {} {\bibfield
  {journal} {\bibinfo  {journal} {Physical Review B}\ }\textbf {\bibinfo
  {volume} {100}},\ \bibinfo {pages} {224104} (\bibinfo {year}
  {2019}{\natexlab{b}})}\BibitemShut {NoStop}%
\bibitem [{\citenamefont {Ojwang}\ \emph {et~al.}(2012)\citenamefont {Ojwang},
  \citenamefont {McWilliams}, \citenamefont {Ke},\ and\ \citenamefont
  {Goncharov}}]{Ojwang2012}%
  \BibitemOpen
  \bibfield  {author} {\bibinfo {author} {\bibfnamefont {J.~G.~O.}\
  \bibnamefont {Ojwang}}, \bibinfo {author} {\bibfnamefont {R.~S.}\
  \bibnamefont {McWilliams}}, \bibinfo {author} {\bibfnamefont
  {X.}~\bibnamefont {Ke}}, \ and\ \bibinfo {author} {\bibfnamefont {A.~F.}\
  \bibnamefont {Goncharov}},\ }\href@noop {} {\bibfield  {journal} {\bibinfo
  {journal} {The Journal of Chemical Physics}\ }\textbf {\bibinfo {volume}
  {137}},\ \bibinfo {pages} {064507} (\bibinfo {year} {2012})}\BibitemShut
  {NoStop}%
\bibitem [{\citenamefont {Palasyuk}\ \emph {et~al.}(2014)\citenamefont
  {Palasyuk}, \citenamefont {Troyan}, \citenamefont {Eremets}, \citenamefont
  {Drozd}, \citenamefont {Medvedev}, \citenamefont {Zaleski-Ejgierd},
  \citenamefont {Magos-Palasyuk}, \citenamefont {Wang}, \citenamefont {Bonev},
  \citenamefont {Dudenko},\ and\ \citenamefont {Naumov}}]{Palasyuk2014}%
  \BibitemOpen
  \bibfield  {author} {\bibinfo {author} {\bibfnamefont {T.}~\bibnamefont
  {Palasyuk}}, \bibinfo {author} {\bibfnamefont {I.}~\bibnamefont {Troyan}},
  \bibinfo {author} {\bibfnamefont {M.}~\bibnamefont {Eremets}}, \bibinfo
  {author} {\bibfnamefont {V.}~\bibnamefont {Drozd}}, \bibinfo {author}
  {\bibfnamefont {S.}~\bibnamefont {Medvedev}}, \bibinfo {author}
  {\bibfnamefont {P.}~\bibnamefont {Zaleski-Ejgierd}}, \bibinfo {author}
  {\bibfnamefont {E.}~\bibnamefont {Magos-Palasyuk}}, \bibinfo {author}
  {\bibfnamefont {H.~B.}\ \bibnamefont {Wang}}, \bibinfo {author}
  {\bibfnamefont {S.~A.}\ \bibnamefont {Bonev}}, \bibinfo {author}
  {\bibfnamefont {D.}~\bibnamefont {Dudenko}}, \ and\ \bibinfo {author}
  {\bibfnamefont {P.}~\bibnamefont {Naumov}},\ }\href@noop {} {\bibfield
  {journal} {\bibinfo  {journal} {Nature Communications}\ }\textbf {\bibinfo
  {volume} {5}},\ \bibinfo {pages} {3460} (\bibinfo {year} {2014})}\BibitemShut
  {NoStop}%
\bibitem [{\citenamefont {Dick}(1981)}]{Dick1981}%
  \BibitemOpen
  \bibfield  {author} {\bibinfo {author} {\bibfnamefont {R.~D.}\ \bibnamefont
  {Dick}},\ }\href {\doibase {10.1063/1.441586}} {\bibfield  {journal}
  {\bibinfo  {journal} {{The Journal of Chemical Physics}}\ }\textbf {\bibinfo
  {volume} {{74}}},\ \bibinfo {pages} {{4053}} (\bibinfo {year}
  {{1981}})}\BibitemShut {NoStop}%
\bibitem [{\citenamefont {Mitchell}\ and\ \citenamefont
  {Nellis}(1982)}]{Mitchell1982}%
  \BibitemOpen
  \bibfield  {author} {\bibinfo {author} {\bibfnamefont {A.~C.}\ \bibnamefont
  {Mitchell}}\ and\ \bibinfo {author} {\bibfnamefont {W.~J.}\ \bibnamefont
  {Nellis}},\ }\href {\doibase {10.1063/1.443030}} {\bibfield  {journal}
  {\bibinfo  {journal} {The Journal of Chemical Physics}\ }\textbf {\bibinfo
  {volume} {{76}}},\ \bibinfo {pages} {{6273}} (\bibinfo {year}
  {{1982}})}\BibitemShut {NoStop}%
\bibitem [{\citenamefont {Nellis}\ \emph {et~al.}(1988)\citenamefont {Nellis},
  \citenamefont {Hamilton}, \citenamefont {Holmes}, \citenamefont {Radousky},
  \citenamefont {Ree}, \citenamefont {Mitchell},\ and\ \citenamefont
  {Nicol}}]{Nellis1988}%
  \BibitemOpen
  \bibfield  {author} {\bibinfo {author} {\bibfnamefont {W.~J.}\ \bibnamefont
  {Nellis}}, \bibinfo {author} {\bibfnamefont {D.~C.}\ \bibnamefont
  {Hamilton}}, \bibinfo {author} {\bibfnamefont {N.~C.}\ \bibnamefont
  {Holmes}}, \bibinfo {author} {\bibfnamefont {H.~B.}\ \bibnamefont
  {Radousky}}, \bibinfo {author} {\bibfnamefont {F.~H.}\ \bibnamefont {Ree}},
  \bibinfo {author} {\bibfnamefont {A.~C.}\ \bibnamefont {Mitchell}}, \ and\
  \bibinfo {author} {\bibfnamefont {M.}~\bibnamefont {Nicol}},\ }\href
  {\doibase {10.1126/science.240.4853.779}} {\bibfield  {journal} {\bibinfo
  {journal} {{Science}}\ }\textbf {\bibinfo {volume} {{240}}},\ \bibinfo
  {pages} {{779}} (\bibinfo {year} {{1988}})}\BibitemShut {NoStop}%
\bibitem [{\citenamefont {Radousky}\ \emph {et~al.}(1990)\citenamefont
  {Radousky}, \citenamefont {Mitchell},\ and\ \citenamefont
  {Nellis}}]{Radousky1990}%
  \BibitemOpen
  \bibfield  {author} {\bibinfo {author} {\bibfnamefont {H.~B.}\ \bibnamefont
  {Radousky}}, \bibinfo {author} {\bibfnamefont {A.~C.}\ \bibnamefont
  {Mitchell}}, \ and\ \bibinfo {author} {\bibfnamefont {W.~J.}\ \bibnamefont
  {Nellis}},\ }\href {\doibase {10.1063/1.459302}} {\bibfield  {journal}
  {\bibinfo  {journal} {The Journal of Chemical Physics}\ }\textbf {\bibinfo
  {volume} {{93}}},\ \bibinfo {pages} {{8235}} (\bibinfo {year}
  {{1990}})}\BibitemShut {NoStop}%
\bibitem [{\citenamefont {Nellis}\ \emph {et~al.}(1997)\citenamefont {Nellis},
  \citenamefont {Holmes}, \citenamefont {Mitchell}, \citenamefont {Hamilton},\
  and\ \citenamefont {Nicol}}]{Nellis1997}%
  \BibitemOpen
  \bibfield  {author} {\bibinfo {author} {\bibfnamefont {W.~J.}\ \bibnamefont
  {Nellis}}, \bibinfo {author} {\bibfnamefont {N.~C.}\ \bibnamefont {Holmes}},
  \bibinfo {author} {\bibfnamefont {A.~C.}\ \bibnamefont {Mitchell}}, \bibinfo
  {author} {\bibfnamefont {D.~C.}\ \bibnamefont {Hamilton}}, \ and\ \bibinfo
  {author} {\bibfnamefont {M.}~\bibnamefont {Nicol}},\ }\href {\doibase
  {10.1063/1.475200}} {\bibfield  {journal} {\bibinfo  {journal} {The Journal
  of Chemical Physics}\ }\textbf {\bibinfo {volume} {{107}}},\ \bibinfo {pages}
  {{9096}} (\bibinfo {year} {{1997}})}\BibitemShut {NoStop}%
\bibitem [{\citenamefont {Dattelbaum}\ \emph {et~al.}(2019)\citenamefont
  {Dattelbaum}, \citenamefont {Lang}, \citenamefont {Goodwin}, \citenamefont
  {Gibson}, \citenamefont {Gammel}, \citenamefont {Coe}, \citenamefont
  {Ticknor},\ and\ \citenamefont {Leiding}}]{Dattelbaum2019}%
  \BibitemOpen
  \bibfield  {author} {\bibinfo {author} {\bibfnamefont {D.~M.}\ \bibnamefont
  {Dattelbaum}}, \bibinfo {author} {\bibfnamefont {J.~M.}\ \bibnamefont
  {Lang}}, \bibinfo {author} {\bibfnamefont {P.~M.}\ \bibnamefont {Goodwin}},
  \bibinfo {author} {\bibfnamefont {L.~L.}\ \bibnamefont {Gibson}}, \bibinfo
  {author} {\bibfnamefont {W.~P.}\ \bibnamefont {Gammel}}, \bibinfo {author}
  {\bibfnamefont {J.~D.}\ \bibnamefont {Coe}}, \bibinfo {author} {\bibfnamefont
  {C.}~\bibnamefont {Ticknor}}, \ and\ \bibinfo {author} {\bibfnamefont
  {J.~A.}\ \bibnamefont {Leiding}},\ }\href@noop {} {\bibfield  {journal}
  {\bibinfo  {journal} {The Journal of Chemical Physics}\ }\textbf {\bibinfo
  {volume} {150}},\ \bibinfo {pages} {024305} (\bibinfo {year}
  {2019})}\BibitemShut {NoStop}%
\bibitem [{\citenamefont {{Barker}}\ and\ \citenamefont
  {{Hollenbach}}(1972)}]{Barker1972}%
  \BibitemOpen
  \bibfield  {author} {\bibinfo {author} {\bibfnamefont {L.~M.}\ \bibnamefont
  {{Barker}}}\ and\ \bibinfo {author} {\bibfnamefont {R.~E.}\ \bibnamefont
  {{Hollenbach}}},\ }\href {\doibase 10.1063/1.1660986} {\bibfield  {journal}
  {\bibinfo  {journal} {Journal of Applied Physics}\ }\textbf {\bibinfo
  {volume} {43}},\ \bibinfo {pages} {4669} (\bibinfo {year}
  {1972})}\BibitemShut {NoStop}%
\bibitem [{\citenamefont {Celliers}\ \emph
  {et~al.}(2004{\natexlab{a}})\citenamefont {Celliers}, \citenamefont
  {Bradley}, \citenamefont {Collins}, \citenamefont {Hicks}, \citenamefont
  {Boehly},\ and\ \citenamefont {Armstrong}}]{Celliers2004_VISAR}%
  \BibitemOpen
  \bibfield  {author} {\bibinfo {author} {\bibfnamefont {P.~M.}\ \bibnamefont
  {Celliers}}, \bibinfo {author} {\bibfnamefont {D.~K.}\ \bibnamefont
  {Bradley}}, \bibinfo {author} {\bibfnamefont {G.~W.}\ \bibnamefont
  {Collins}}, \bibinfo {author} {\bibfnamefont {D.~G.}\ \bibnamefont {Hicks}},
  \bibinfo {author} {\bibfnamefont {T.~R.}\ \bibnamefont {Boehly}}, \ and\
  \bibinfo {author} {\bibfnamefont {W.~J.}\ \bibnamefont {Armstrong}},\ }\href
  {\doibase 10.1063/1.1807008} {\bibfield  {journal} {\bibinfo  {journal}
  {Review of Scientific Instruments}\ }\textbf {\bibinfo {volume} {75}},\
  \bibinfo {pages} {4916} (\bibinfo {year} {2004}{\natexlab{a}})}\BibitemShut
  {NoStop}%
\bibitem [{\citenamefont {Holmes}(1995)}]{Holmes1995}%
  \BibitemOpen
  \bibfield  {author} {\bibinfo {author} {\bibfnamefont {N.~C.}\ \bibnamefont
  {Holmes}},\ }\href {\doibase 10.1063/1.1145597} {\bibfield  {journal}
  {\bibinfo  {journal} {Review of Scientific Instruments}\ }\textbf {\bibinfo
  {volume} {66}},\ \bibinfo {pages} {2615} (\bibinfo {year}
  {1995})}\BibitemShut {NoStop}%
\bibitem [{\citenamefont {Hall}\ \emph {et~al.}(1997)\citenamefont {Hall},
  \citenamefont {Benuzzi}, \citenamefont {Batani}, \citenamefont {Beretta},
  \citenamefont {Bossi}, \citenamefont {Faral}, \citenamefont {Koenig},
  \citenamefont {Krishnan}, \citenamefont {L\"ower},\ and\ \citenamefont
  {Mahdieh}}]{Hall1997}%
  \BibitemOpen
  \bibfield  {author} {\bibinfo {author} {\bibfnamefont {T.~A.}\ \bibnamefont
  {Hall}}, \bibinfo {author} {\bibfnamefont {A.}~\bibnamefont {Benuzzi}},
  \bibinfo {author} {\bibfnamefont {D.}~\bibnamefont {Batani}}, \bibinfo
  {author} {\bibfnamefont {D.}~\bibnamefont {Beretta}}, \bibinfo {author}
  {\bibfnamefont {S.}~\bibnamefont {Bossi}}, \bibinfo {author} {\bibfnamefont
  {B.}~\bibnamefont {Faral}}, \bibinfo {author} {\bibfnamefont
  {M.}~\bibnamefont {Koenig}}, \bibinfo {author} {\bibfnamefont
  {J.}~\bibnamefont {Krishnan}}, \bibinfo {author} {\bibfnamefont
  {T.}~\bibnamefont {L\"ower}}, \ and\ \bibinfo {author} {\bibfnamefont
  {M.}~\bibnamefont {Mahdieh}},\ }\href {\doibase 10.1103/PhysRevE.55.R6356}
  {\bibfield  {journal} {\bibinfo  {journal} {Phys. Rev. E}\ }\textbf {\bibinfo
  {volume} {55}},\ \bibinfo {pages} {R6356} (\bibinfo {year}
  {1997})}\BibitemShut {NoStop}%
\bibitem [{\citenamefont {{Miller}}\ \emph {et~al.}(2007)\citenamefont
  {{Miller}}, \citenamefont {{Boehly}}, \citenamefont {{Melchior}},
  \citenamefont {{Meyerhofer}}, \citenamefont {{Celliers}}, \citenamefont
  {{Eggert}}, \citenamefont {{Hicks}}, \citenamefont {{Sorce}}, \citenamefont
  {{Oertel}},\ and\ \citenamefont {{Emmel}}}]{Miller2007}%
  \BibitemOpen
  \bibfield  {author} {\bibinfo {author} {\bibfnamefont {J.~E.}\ \bibnamefont
  {{Miller}}}, \bibinfo {author} {\bibfnamefont {T.~R.}\ \bibnamefont
  {{Boehly}}}, \bibinfo {author} {\bibfnamefont {A.}~\bibnamefont
  {{Melchior}}}, \bibinfo {author} {\bibfnamefont {D.~D.}\ \bibnamefont
  {{Meyerhofer}}}, \bibinfo {author} {\bibfnamefont {P.~M.}\ \bibnamefont
  {{Celliers}}}, \bibinfo {author} {\bibfnamefont {J.~H.}\ \bibnamefont
  {{Eggert}}}, \bibinfo {author} {\bibfnamefont {D.~G.}\ \bibnamefont
  {{Hicks}}}, \bibinfo {author} {\bibfnamefont {C.~M.}\ \bibnamefont
  {{Sorce}}}, \bibinfo {author} {\bibfnamefont {J.~A.}\ \bibnamefont
  {{Oertel}}}, \ and\ \bibinfo {author} {\bibfnamefont {P.~M.}\ \bibnamefont
  {{Emmel}}},\ }\href@noop {} {\bibfield  {journal} {\bibinfo  {journal}
  {Review of Scientific Instruments}\ }\textbf {\bibinfo {volume} {78}},\
  \bibinfo {eid} {034903-034903-7} (\bibinfo {year} {2007})}\BibitemShut
  {NoStop}%
\bibitem [{\citenamefont {Forbes}(2012)}]{forbes}%
  \BibitemOpen
  \bibfield  {author} {\bibinfo {author} {\bibfnamefont {J.~W.}\ \bibnamefont
  {Forbes}},\ }\href@noop {} {\emph {\bibinfo {title} {Shock Wave Compression
  of Condensed Matter}}}\ (\bibinfo  {publisher} {Springer},\ \bibinfo {year}
  {2012})\BibitemShut {NoStop}%
\bibitem [{\citenamefont {Brygoo}\ \emph {et~al.}(2015)\citenamefont {Brygoo},
  \citenamefont {Millot}, \citenamefont {Loubeyre}, \citenamefont {Lazicki},
  \citenamefont {Hamel}, \citenamefont {Qi}, \citenamefont {Celliers},
  \citenamefont {Coppari}, \citenamefont {Eggert}, \citenamefont {Fratanduono}
  \emph {et~al.}}]{Brygoo2015}%
  \BibitemOpen
  \bibfield  {author} {\bibinfo {author} {\bibfnamefont {S.}~\bibnamefont
  {Brygoo}}, \bibinfo {author} {\bibfnamefont {M.}~\bibnamefont {Millot}},
  \bibinfo {author} {\bibfnamefont {P.}~\bibnamefont {Loubeyre}}, \bibinfo
  {author} {\bibfnamefont {A.~E.}\ \bibnamefont {Lazicki}}, \bibinfo {author}
  {\bibfnamefont {S.}~\bibnamefont {Hamel}}, \bibinfo {author} {\bibfnamefont
  {T.}~\bibnamefont {Qi}}, \bibinfo {author} {\bibfnamefont {P.~M.}\
  \bibnamefont {Celliers}}, \bibinfo {author} {\bibfnamefont {F.}~\bibnamefont
  {Coppari}}, \bibinfo {author} {\bibfnamefont {J.~H.}\ \bibnamefont {Eggert}},
  \bibinfo {author} {\bibfnamefont {D.~E.}\ \bibnamefont {Fratanduono}},  \emph
  {et~al.},\ }\href@noop {} {\bibfield  {journal} {\bibinfo  {journal} {Journal
  of Applied Physics}\ }\textbf {\bibinfo {volume} {118}},\ \bibinfo {pages}
  {195901} (\bibinfo {year} {2015})}\BibitemShut {NoStop}%
\bibitem [{\citenamefont {Zel'dovich}\ and\ \citenamefont
  {Raizer}(2002)}]{zeldovich}%
  \BibitemOpen
  \bibfield  {author} {\bibinfo {author} {\bibfnamefont {Y.~B.}\ \bibnamefont
  {Zel'dovich}}\ and\ \bibinfo {author} {\bibfnamefont {Y.~P.}\ \bibnamefont
  {Raizer}},\ }\href@noop {} {\emph {\bibinfo {title} {Physics of shock waves
  and high-temperature hydrodynamic phenomena}}}\ (\bibinfo  {publisher} {Dover
  Publications},\ \bibinfo {address} {New York},\ \bibinfo {year} {2002})\ pp.\
  \bibinfo {pages} {XXVII, 916 S.}\BibitemShut {Stop}%
\bibitem [{Note1()}]{Note1}%
  \BibitemOpen
  \bibinfo {note} {See Supplementary Material for further experimental details,
  which includes Refs.~\cite {Ghosh1999, Gauthier1988, Kume1998, Robertson1973,
  Tillner1993, Huser2015, Qi2015, Gregor2016}}\BibitemShut {NoStop}%
\bibitem [{\citenamefont {Kresse}\ and\ \citenamefont
  {Hafner}(1993{\natexlab{a}})}]{Kresse1993a}%
  \BibitemOpen
  \bibfield  {author} {\bibinfo {author} {\bibfnamefont {G.}~\bibnamefont
  {Kresse}}\ and\ \bibinfo {author} {\bibfnamefont {J.}~\bibnamefont
  {Hafner}},\ }\href {\doibase {10.1103/PhysRevB.47.558}} {\bibfield  {journal}
  {\bibinfo  {journal} {{Physical Review B}}\ }\textbf {\bibinfo {volume}
  {{47}}},\ \bibinfo {pages} {{558}} (\bibinfo {year}
  {{1993}}{\natexlab{a}})}\BibitemShut {NoStop}%
\bibitem [{\citenamefont {Kresse}\ and\ \citenamefont
  {Hafner}(1993{\natexlab{b}})}]{Kresse1993b}%
  \BibitemOpen
  \bibfield  {author} {\bibinfo {author} {\bibfnamefont {G.}~\bibnamefont
  {Kresse}}\ and\ \bibinfo {author} {\bibfnamefont {J.}~\bibnamefont
  {Hafner}},\ }\href@noop {} {\bibfield  {journal} {\bibinfo  {journal}
  {{Physical Review B}}\ }\textbf {\bibinfo {volume} {{48}}},\ \bibinfo {pages}
  {{13115}} (\bibinfo {year} {{1993}}{\natexlab{b}})}\BibitemShut {NoStop}%
\bibitem [{\citenamefont {{Kresse}}\ and\ \citenamefont
  {{Hafner}}(1994)}]{Kresse1994}%
  \BibitemOpen
  \bibfield  {author} {\bibinfo {author} {\bibfnamefont {G.}~\bibnamefont
  {{Kresse}}}\ and\ \bibinfo {author} {\bibfnamefont {J.}~\bibnamefont
  {{Hafner}}},\ }\href {\doibase {10.1103/PhysRevB.49.14251}} {\bibfield
  {journal} {\bibinfo  {journal} {{Physical Review B}}\ }\textbf {\bibinfo
  {volume} {{49}}},\ \bibinfo {pages} {{14251}} (\bibinfo {year}
  {{1994}})}\BibitemShut {NoStop}%
\bibitem [{\citenamefont {Kresse}\ and\ \citenamefont
  {Furthm\"uller}(1996)}]{Kresse1996}%
  \BibitemOpen
  \bibfield  {author} {\bibinfo {author} {\bibfnamefont {G.}~\bibnamefont
  {Kresse}}\ and\ \bibinfo {author} {\bibfnamefont {J.}~\bibnamefont
  {Furthm\"uller}},\ }\href {\doibase {10.1103/PhysRevB.54.11169}} {\bibfield
  {journal} {\bibinfo  {journal} {Physical Review B}\ }\textbf {\bibinfo
  {volume} {{54}}},\ \bibinfo {pages} {{11169}} (\bibinfo {year}
  {{1996}})}\BibitemShut {NoStop}%
\bibitem [{\citenamefont {Perdew}\ \emph {et~al.}(1996)\citenamefont {Perdew},
  \citenamefont {Burke},\ and\ \citenamefont {Ernzerhof}}]{Perdew1996}%
  \BibitemOpen
  \bibfield  {author} {\bibinfo {author} {\bibfnamefont {J.~P.}\ \bibnamefont
  {Perdew}}, \bibinfo {author} {\bibfnamefont {K.}~\bibnamefont {Burke}}, \
  and\ \bibinfo {author} {\bibfnamefont {M.}~\bibnamefont {Ernzerhof}},\ }\href
  {\doibase {10.1103/PhysRevLett.77.3865}} {\bibfield  {journal} {\bibinfo
  {journal} {{Physical Review Letters}}\ }\textbf {\bibinfo {volume} {{77}}},\
  \bibinfo {pages} {{3865}} (\bibinfo {year} {{1996}})}\BibitemShut {NoStop}%
\bibitem [{\citenamefont {Heyd}\ \emph {et~al.}(2003)\citenamefont {Heyd},
  \citenamefont {Scuseria},\ and\ \citenamefont {Ernzerhof}}]{Heyd2003}%
  \BibitemOpen
  \bibfield  {author} {\bibinfo {author} {\bibfnamefont {J.}~\bibnamefont
  {Heyd}}, \bibinfo {author} {\bibfnamefont {G.~E.}\ \bibnamefont {Scuseria}},
  \ and\ \bibinfo {author} {\bibfnamefont {M.}~\bibnamefont {Ernzerhof}},\
  }\href {\doibase {10.1063/1.1564060}} {\bibfield  {journal} {\bibinfo
  {journal} {The Journal of Chemical Physics}\ }\textbf {\bibinfo {volume}
  {{118}}},\ \bibinfo {pages} {{8207}} (\bibinfo {year} {{2003}})}\BibitemShut
  {NoStop}%
\bibitem [{\citenamefont {Heyd}\ \emph {et~al.}(2006)\citenamefont {Heyd},
  \citenamefont {Scuseria},\ and\ \citenamefont {Ernzerhof}}]{Heyd2006}%
  \BibitemOpen
  \bibfield  {author} {\bibinfo {author} {\bibfnamefont {J.}~\bibnamefont
  {Heyd}}, \bibinfo {author} {\bibfnamefont {G.~E.}\ \bibnamefont {Scuseria}},
  \ and\ \bibinfo {author} {\bibfnamefont {M.}~\bibnamefont {Ernzerhof}},\
  }\href@noop {} {\bibfield  {journal} {\bibinfo  {journal} {J. Chem. Phys.}\
  }\textbf {\bibinfo {volume} {124}},\ \bibinfo {pages} {219906} (\bibinfo
  {year} {2006})}\BibitemShut {NoStop}%
\bibitem [{Note2()}]{Note2}%
  \BibitemOpen
  \bibinfo {note} {See Supplementary Material for further computational
  details, which includes Refs.~\cite {Nose1984, Baldereschi1973, Kubo1957,
  Greenwood1958, Holst2011, Gajdos2006, Berens1983}}\BibitemShut {NoStop}%
\bibitem [{\citenamefont {Kovel}(1973)}]{Kovel1973}%
  \BibitemOpen
  \bibfield  {author} {\bibinfo {author} {\bibfnamefont {M.~I.}\ \bibnamefont
  {Kovel}},\ }\emph {\bibinfo {title} {The Shock Wave Hugoniot and Electrical
  Conductivity of Liquid Ammonia in the Pressure Range 45 Kb to 282 Kb}},\
  \href@noop {} {Ph.D. thesis},\ \bibinfo  {school} {University of California}
  (\bibinfo {year} {1973})\BibitemShut {NoStop}%
\bibitem [{\citenamefont {{Millot}}\ \emph {et~al.}(2015)\citenamefont
  {{Millot}}, \citenamefont {{Dubrovinskaia}}, \citenamefont {{{\v C}ernok}},
  \citenamefont {{Blaha}}, \citenamefont {{Dubrovinsky}}, \citenamefont
  {{Braun}}, \citenamefont {{Celliers}}, \citenamefont {{Collins}},
  \citenamefont {{Eggert}},\ and\ \citenamefont {{Jeanloz}}}]{Millot2015}%
  \BibitemOpen
  \bibfield  {author} {\bibinfo {author} {\bibfnamefont {M.}~\bibnamefont
  {{Millot}}}, \bibinfo {author} {\bibfnamefont {N.}~\bibnamefont
  {{Dubrovinskaia}}}, \bibinfo {author} {\bibfnamefont {A.}~\bibnamefont {{{\v
  C}ernok}}}, \bibinfo {author} {\bibfnamefont {S.}~\bibnamefont {{Blaha}}},
  \bibinfo {author} {\bibfnamefont {L.}~\bibnamefont {{Dubrovinsky}}}, \bibinfo
  {author} {\bibfnamefont {D.~G.}\ \bibnamefont {{Braun}}}, \bibinfo {author}
  {\bibfnamefont {P.~M.}\ \bibnamefont {{Celliers}}}, \bibinfo {author}
  {\bibfnamefont {G.~W.}\ \bibnamefont {{Collins}}}, \bibinfo {author}
  {\bibfnamefont {J.~H.}\ \bibnamefont {{Eggert}}}, \ and\ \bibinfo {author}
  {\bibfnamefont {R.}~\bibnamefont {{Jeanloz}}},\ }\href@noop {} {\bibfield
  {journal} {\bibinfo  {journal} {Science}\ }\textbf {\bibinfo {volume}
  {347}},\ \bibinfo {pages} {418} (\bibinfo {year} {2015})}\BibitemShut
  {NoStop}%
\bibitem [{\citenamefont {McWilliams}\ \emph {et~al.}(2012)\citenamefont
  {McWilliams}, \citenamefont {Spaulding}, \citenamefont {Eggert},
  \citenamefont {Celliers}, \citenamefont {Hicks}, \citenamefont {Smith},
  \citenamefont {Collins},\ and\ \citenamefont {Jeanloz}}]{McWilliams2012}%
  \BibitemOpen
  \bibfield  {author} {\bibinfo {author} {\bibfnamefont {R.~S.}\ \bibnamefont
  {McWilliams}}, \bibinfo {author} {\bibfnamefont {D.~K.}\ \bibnamefont
  {Spaulding}}, \bibinfo {author} {\bibfnamefont {J.~H.}\ \bibnamefont
  {Eggert}}, \bibinfo {author} {\bibfnamefont {P.~M.}\ \bibnamefont
  {Celliers}}, \bibinfo {author} {\bibfnamefont {D.~G.}\ \bibnamefont {Hicks}},
  \bibinfo {author} {\bibfnamefont {R.~F.}\ \bibnamefont {Smith}}, \bibinfo
  {author} {\bibfnamefont {G.~W.}\ \bibnamefont {Collins}}, \ and\ \bibinfo
  {author} {\bibfnamefont {R.}~\bibnamefont {Jeanloz}},\ }\href@noop {}
  {\bibfield  {journal} {\bibinfo  {journal} {Science}\ }\textbf {\bibinfo
  {volume} {338}},\ \bibinfo {pages} {1330} (\bibinfo {year}
  {2012})}\BibitemShut {NoStop}%
\bibitem [{\citenamefont {French}\ \emph {et~al.}(2010)\citenamefont {French},
  \citenamefont {Mattsson},\ and\ \citenamefont {Redmer}}]{French2010}%
  \BibitemOpen
  \bibfield  {author} {\bibinfo {author} {\bibfnamefont {M.}~\bibnamefont
  {French}}, \bibinfo {author} {\bibfnamefont {T.~R.}\ \bibnamefont
  {Mattsson}}, \ and\ \bibinfo {author} {\bibfnamefont {R.}~\bibnamefont
  {Redmer}},\ }\href {\doibase {10.1103/PhysRevB.82.174108}} {\bibfield
  {journal} {\bibinfo  {journal} {{Physical Review B}}\ }\textbf {\bibinfo
  {volume} {{82}}},\ \bibinfo {pages} {174108} (\bibinfo {year}
  {{2010}})}\BibitemShut {NoStop}%
\bibitem [{\citenamefont {Celliers}\ \emph
  {et~al.}(2004{\natexlab{b}})\citenamefont {Celliers}, \citenamefont
  {Collins}, \citenamefont {Hicks}, \citenamefont {Koenig}, \citenamefont
  {Henry}, \citenamefont {Benuzzi-Mounaix}, \citenamefont {Batani},
  \citenamefont {Bradley}, \citenamefont {Silva}, \citenamefont {Wallace},
  \citenamefont {Moon}, \citenamefont {Eggert}, \citenamefont {Lee},
  \citenamefont {Benedetti}, \citenamefont {Jeanloz}, \citenamefont {Masclet},
  \citenamefont {Dague}, \citenamefont {Marchet}, \citenamefont {Gloahec},
  \citenamefont {Reverdin}, \citenamefont {Pasley}, \citenamefont {Willi},
  \citenamefont {Neely},\ and\ \citenamefont {Danson}}]{Celliers2004}%
  \BibitemOpen
  \bibfield  {author} {\bibinfo {author} {\bibfnamefont {P.~M.}\ \bibnamefont
  {Celliers}}, \bibinfo {author} {\bibfnamefont {G.~W.}\ \bibnamefont
  {Collins}}, \bibinfo {author} {\bibfnamefont {D.~G.}\ \bibnamefont {Hicks}},
  \bibinfo {author} {\bibfnamefont {M.}~\bibnamefont {Koenig}}, \bibinfo
  {author} {\bibfnamefont {E.}~\bibnamefont {Henry}}, \bibinfo {author}
  {\bibfnamefont {A.}~\bibnamefont {Benuzzi-Mounaix}}, \bibinfo {author}
  {\bibfnamefont {D.}~\bibnamefont {Batani}}, \bibinfo {author} {\bibfnamefont
  {D.~K.}\ \bibnamefont {Bradley}}, \bibinfo {author} {\bibfnamefont
  {L.~B.~D.}\ \bibnamefont {Silva}}, \bibinfo {author} {\bibfnamefont {R.~J.}\
  \bibnamefont {Wallace}}, \bibinfo {author} {\bibfnamefont {S.~J.}\
  \bibnamefont {Moon}}, \bibinfo {author} {\bibfnamefont {J.~H.}\ \bibnamefont
  {Eggert}}, \bibinfo {author} {\bibfnamefont {K.~K.~M.}\ \bibnamefont {Lee}},
  \bibinfo {author} {\bibfnamefont {L.~R.}\ \bibnamefont {Benedetti}}, \bibinfo
  {author} {\bibfnamefont {R.}~\bibnamefont {Jeanloz}}, \bibinfo {author}
  {\bibfnamefont {I.}~\bibnamefont {Masclet}}, \bibinfo {author} {\bibfnamefont
  {N.}~\bibnamefont {Dague}}, \bibinfo {author} {\bibfnamefont
  {B.}~\bibnamefont {Marchet}}, \bibinfo {author} {\bibfnamefont {M.~R.~L.}\
  \bibnamefont {Gloahec}}, \bibinfo {author} {\bibfnamefont {C.}~\bibnamefont
  {Reverdin}}, \bibinfo {author} {\bibfnamefont {J.}~\bibnamefont {Pasley}},
  \bibinfo {author} {\bibfnamefont {O.}~\bibnamefont {Willi}}, \bibinfo
  {author} {\bibfnamefont {D.}~\bibnamefont {Neely}}, \ and\ \bibinfo {author}
  {\bibfnamefont {C.}~\bibnamefont {Danson}},\ }\href@noop {} {\bibfield
  {journal} {\bibinfo  {journal} {Physics of Plasmas}\ }\textbf {\bibinfo
  {volume} {11}},\ \bibinfo {pages} {L41} (\bibinfo {year}
  {2004}{\natexlab{b}})}\BibitemShut {NoStop}%
\bibitem [{\citenamefont {Knudson}\ \emph {et~al.}(2012)\citenamefont
  {Knudson}, \citenamefont {Desjarlais}, \citenamefont {Lemke}, \citenamefont
  {Mattsson}, \citenamefont {French}, \citenamefont {Nettelmann},\ and\
  \citenamefont {Redmer}}]{Knudson2012}%
  \BibitemOpen
  \bibfield  {author} {\bibinfo {author} {\bibfnamefont {M.~D.}\ \bibnamefont
  {Knudson}}, \bibinfo {author} {\bibfnamefont {M.~P.}\ \bibnamefont
  {Desjarlais}}, \bibinfo {author} {\bibfnamefont {R.~W.}\ \bibnamefont
  {Lemke}}, \bibinfo {author} {\bibfnamefont {T.~R.}\ \bibnamefont {Mattsson}},
  \bibinfo {author} {\bibfnamefont {M.}~\bibnamefont {French}}, \bibinfo
  {author} {\bibfnamefont {N.}~\bibnamefont {Nettelmann}}, \ and\ \bibinfo
  {author} {\bibfnamefont {R.}~\bibnamefont {Redmer}},\ }\href@noop {}
  {\bibfield  {journal} {\bibinfo  {journal} {{Physical Review Letters}}\
  }\textbf {\bibinfo {volume} {{108}}},\ \bibinfo {pages} {091102} (\bibinfo
  {year} {{2012}})}\BibitemShut {NoStop}%
\bibitem [{\citenamefont {Kimura}\ \emph {et~al.}(2015)\citenamefont {Kimura},
  \citenamefont {Ozaki}, \citenamefont {Sano}, \citenamefont {Okuchi},
  \citenamefont {Sano}, \citenamefont {Shimizu}, \citenamefont {Miyanishi},
  \citenamefont {Terai}, \citenamefont {Kakeshita}, \citenamefont {Sakawa},\
  and\ \citenamefont {Kodama}}]{Kimura2015}%
  \BibitemOpen
  \bibfield  {author} {\bibinfo {author} {\bibfnamefont {T.}~\bibnamefont
  {Kimura}}, \bibinfo {author} {\bibfnamefont {N.}~\bibnamefont {Ozaki}},
  \bibinfo {author} {\bibfnamefont {T.}~\bibnamefont {Sano}}, \bibinfo {author}
  {\bibfnamefont {T.}~\bibnamefont {Okuchi}}, \bibinfo {author} {\bibfnamefont
  {T.}~\bibnamefont {Sano}}, \bibinfo {author} {\bibfnamefont {K.}~\bibnamefont
  {Shimizu}}, \bibinfo {author} {\bibfnamefont {K.}~\bibnamefont {Miyanishi}},
  \bibinfo {author} {\bibfnamefont {T.}~\bibnamefont {Terai}}, \bibinfo
  {author} {\bibfnamefont {T.}~\bibnamefont {Kakeshita}}, \bibinfo {author}
  {\bibfnamefont {Y.}~\bibnamefont {Sakawa}}, \ and\ \bibinfo {author}
  {\bibfnamefont {R.}~\bibnamefont {Kodama}},\ }\href@noop {} {\bibfield
  {journal} {\bibinfo  {journal} {Journal of Chemical Physics}\ }\textbf
  {\bibinfo {volume} {142}},\ \bibinfo {pages} {164504} (\bibinfo {year}
  {2015})}\BibitemShut {NoStop}%
\bibitem [{\citenamefont {Millot}\ \emph {et~al.}(2018)\citenamefont {Millot},
  \citenamefont {Hamel}, \citenamefont {Rygg}, \citenamefont {Celliers},
  \citenamefont {Collins}, \citenamefont {Coppari}, \citenamefont
  {Fratanduono}, \citenamefont {Jeanloz}, \citenamefont {Swift},\ and\
  \citenamefont {Eggert}}]{Millot2018}%
  \BibitemOpen
  \bibfield  {author} {\bibinfo {author} {\bibfnamefont {M.}~\bibnamefont
  {Millot}}, \bibinfo {author} {\bibfnamefont {S.}~\bibnamefont {Hamel}},
  \bibinfo {author} {\bibfnamefont {J.~R.}\ \bibnamefont {Rygg}}, \bibinfo
  {author} {\bibfnamefont {P.~M.}\ \bibnamefont {Celliers}}, \bibinfo {author}
  {\bibfnamefont {G.~W.}\ \bibnamefont {Collins}}, \bibinfo {author}
  {\bibfnamefont {F.}~\bibnamefont {Coppari}}, \bibinfo {author} {\bibfnamefont
  {D.~E.}\ \bibnamefont {Fratanduono}}, \bibinfo {author} {\bibfnamefont
  {R.}~\bibnamefont {Jeanloz}}, \bibinfo {author} {\bibfnamefont {D.~C.}\
  \bibnamefont {Swift}}, \ and\ \bibinfo {author} {\bibfnamefont {J.~H.}\
  \bibnamefont {Eggert}},\ }\href@noop {} {\bibfield  {journal} {\bibinfo
  {journal} {Nature Physics}\ }\textbf {\bibinfo {volume} {14}},\ \bibinfo
  {pages} {297} (\bibinfo {year} {2018})}\BibitemShut {NoStop}%
\bibitem [{\citenamefont {Naden~Robinson}\ \emph {et~al.}(2017)\citenamefont
  {Naden~Robinson}, \citenamefont {Wang}, \citenamefont {Ma},\ and\
  \citenamefont {Hermann}}]{NadenRobinson2017}%
  \BibitemOpen
  \bibfield  {author} {\bibinfo {author} {\bibfnamefont {V.}~\bibnamefont
  {Naden~Robinson}}, \bibinfo {author} {\bibfnamefont {Y.~C.}\ \bibnamefont
  {Wang}}, \bibinfo {author} {\bibfnamefont {Y.~M.}\ \bibnamefont {Ma}}, \ and\
  \bibinfo {author} {\bibfnamefont {A.}~\bibnamefont {Hermann}},\ }\href
  {\doibase 10.1073/pnas.1706244114} {\bibfield  {journal} {\bibinfo  {journal}
  {PNAS}\ }\textbf {\bibinfo {volume} {114}},\ \bibinfo {pages} {9003}
  (\bibinfo {year} {2017})}\BibitemShut {NoStop}%
\bibitem [{\citenamefont {Ghosh}(1999)}]{Ghosh1999}%
  \BibitemOpen
  \bibfield  {author} {\bibinfo {author} {\bibfnamefont {G.}~\bibnamefont
  {Ghosh}},\ }\href@noop {} {\bibfield  {journal} {\bibinfo  {journal} {Optics
  communications}\ }\textbf {\bibinfo {volume} {163}},\ \bibinfo {pages} {95}
  (\bibinfo {year} {1999})}\BibitemShut {NoStop}%
\bibitem [{\citenamefont {Gauthier}\ \emph {et~al.}(1988)\citenamefont
  {Gauthier}, \citenamefont {Pruzan}, \citenamefont {Chervin},\ and\
  \citenamefont {Polian}}]{Gauthier1988}%
  \BibitemOpen
  \bibfield  {author} {\bibinfo {author} {\bibfnamefont {M.}~\bibnamefont
  {Gauthier}}, \bibinfo {author} {\bibfnamefont {P.}~\bibnamefont {Pruzan}},
  \bibinfo {author} {\bibfnamefont {J.~C.}\ \bibnamefont {Chervin}}, \ and\
  \bibinfo {author} {\bibfnamefont {A.}~\bibnamefont {Polian}},\ }\href
  {\doibase {10.1016/0038-1098(88)90263-3}} {\bibfield  {journal} {\bibinfo
  {journal} {{Solid State Communications}}\ }\textbf {\bibinfo {volume}
  {{68}}},\ \bibinfo {pages} {{149}} (\bibinfo {year} {{1988}})}\BibitemShut
  {NoStop}%
\bibitem [{\citenamefont {Kume}\ \emph {et~al.}(1998)\citenamefont {Kume},
  \citenamefont {Daimon}, \citenamefont {Sasaki},\ and\ \citenamefont
  {Shimizu}}]{Kume1998}%
  \BibitemOpen
  \bibfield  {author} {\bibinfo {author} {\bibfnamefont {T.}~\bibnamefont
  {Kume}}, \bibinfo {author} {\bibfnamefont {M.}~\bibnamefont {Daimon}},
  \bibinfo {author} {\bibfnamefont {S.}~\bibnamefont {Sasaki}}, \ and\ \bibinfo
  {author} {\bibfnamefont {H.}~\bibnamefont {Shimizu}},\ }\href {\doibase
  {10.1103/PhysRevB.57.13347}} {\bibfield  {journal} {\bibinfo  {journal}
  {{Physical Review B}}\ }\textbf {\bibinfo {volume} {{57}}},\ \bibinfo {pages}
  {{13347}} (\bibinfo {year} {{1998}})}\BibitemShut {NoStop}%
\bibitem [{\citenamefont {Robertson}\ and\ \citenamefont
  {Williams}(1973)}]{Robertson1973}%
  \BibitemOpen
  \bibfield  {author} {\bibinfo {author} {\bibfnamefont {C.~W.}\ \bibnamefont
  {Robertson}}\ and\ \bibinfo {author} {\bibfnamefont {D.}~\bibnamefont
  {Williams}},\ }\href@noop {} {\bibfield  {journal} {\bibinfo  {journal}
  {Journal of the Optical Society of America}\ }\textbf {\bibinfo {volume}
  {63}},\ \bibinfo {pages} {188} (\bibinfo {year} {1973})}\BibitemShut
  {NoStop}%
\bibitem [{\citenamefont {Tillner-Roth}\ \emph {et~al.}(1993)\citenamefont
  {Tillner-Roth}, \citenamefont {Harms-Watzenberg},\ and\ \citenamefont
  {Baehr}}]{Tillner1993}%
  \BibitemOpen
  \bibfield  {author} {\bibinfo {author} {\bibfnamefont {R.}~\bibnamefont
  {Tillner-Roth}}, \bibinfo {author} {\bibfnamefont {F.}~\bibnamefont
  {Harms-Watzenberg}}, \ and\ \bibinfo {author} {\bibfnamefont
  {H.}~\bibnamefont {Baehr}},\ }\href@noop {} {\bibfield  {journal} {\bibinfo
  {journal} {DKV TAGUNGSBERICHT}\ }\textbf {\bibinfo {volume} {20}},\ \bibinfo
  {pages} {67} (\bibinfo {year} {1993})}\BibitemShut {NoStop}%
\bibitem [{\citenamefont {Huser}\ \emph {et~al.}(2015)\citenamefont {Huser},
  \citenamefont {Recoules}, \citenamefont {Ozaki}, \citenamefont {Sano},
  \citenamefont {Sakawa}, \citenamefont {Salin}, \citenamefont {Albertazzi},
  \citenamefont {Miyanishi},\ and\ \citenamefont {Kodama}}]{Huser2015}%
  \BibitemOpen
  \bibfield  {author} {\bibinfo {author} {\bibfnamefont {G.}~\bibnamefont
  {Huser}}, \bibinfo {author} {\bibfnamefont {V.}~\bibnamefont {Recoules}},
  \bibinfo {author} {\bibfnamefont {N.}~\bibnamefont {Ozaki}}, \bibinfo
  {author} {\bibfnamefont {T.}~\bibnamefont {Sano}}, \bibinfo {author}
  {\bibfnamefont {Y.}~\bibnamefont {Sakawa}}, \bibinfo {author} {\bibfnamefont
  {G.}~\bibnamefont {Salin}}, \bibinfo {author} {\bibfnamefont
  {B.}~\bibnamefont {Albertazzi}}, \bibinfo {author} {\bibfnamefont
  {K.}~\bibnamefont {Miyanishi}}, \ and\ \bibinfo {author} {\bibfnamefont
  {R.}~\bibnamefont {Kodama}},\ }\href@noop {} {\bibfield  {journal} {\bibinfo
  {journal} {Physical Review E}\ }\textbf {\bibinfo {volume} {92}},\ \bibinfo
  {pages} {063108} (\bibinfo {year} {2015})}\BibitemShut {NoStop}%
\bibitem [{\citenamefont {Qi}\ \emph {et~al.}(2015)\citenamefont {Qi},
  \citenamefont {Millot}, \citenamefont {Kraus}, \citenamefont {Root},\ and\
  \citenamefont {Hamel}}]{Qi2015}%
  \BibitemOpen
  \bibfield  {author} {\bibinfo {author} {\bibfnamefont {T.}~\bibnamefont
  {Qi}}, \bibinfo {author} {\bibfnamefont {M.}~\bibnamefont {Millot}}, \bibinfo
  {author} {\bibfnamefont {R.~G.}\ \bibnamefont {Kraus}}, \bibinfo {author}
  {\bibfnamefont {S.}~\bibnamefont {Root}}, \ and\ \bibinfo {author}
  {\bibfnamefont {S.}~\bibnamefont {Hamel}},\ }\href@noop {} {\bibfield
  {journal} {\bibinfo  {journal} {Physics of Plasmas}\ }\textbf {\bibinfo
  {volume} {22}},\ \bibinfo {pages} {062706} (\bibinfo {year}
  {2015})}\BibitemShut {NoStop}%
\bibitem [{\citenamefont {Gregor}\ \emph {et~al.}(2016)\citenamefont {Gregor},
  \citenamefont {Boni}, \citenamefont {Sorce}, \citenamefont {Kendrick},
  \citenamefont {McCoy}, \citenamefont {Polsin}, \citenamefont {Boehly},
  \citenamefont {Celliers}, \citenamefont {Collins}, \citenamefont
  {Fratanduono} \emph {et~al.}}]{Gregor2016}%
  \BibitemOpen
  \bibfield  {author} {\bibinfo {author} {\bibfnamefont {M.}~\bibnamefont
  {Gregor}}, \bibinfo {author} {\bibfnamefont {R.}~\bibnamefont {Boni}},
  \bibinfo {author} {\bibfnamefont {A.}~\bibnamefont {Sorce}}, \bibinfo
  {author} {\bibfnamefont {J.}~\bibnamefont {Kendrick}}, \bibinfo {author}
  {\bibfnamefont {C.}~\bibnamefont {McCoy}}, \bibinfo {author} {\bibfnamefont
  {D.}~\bibnamefont {Polsin}}, \bibinfo {author} {\bibfnamefont
  {T.}~\bibnamefont {Boehly}}, \bibinfo {author} {\bibfnamefont
  {P.}~\bibnamefont {Celliers}}, \bibinfo {author} {\bibfnamefont
  {G.}~\bibnamefont {Collins}}, \bibinfo {author} {\bibfnamefont
  {D.}~\bibnamefont {Fratanduono}},  \emph {et~al.},\ }\href@noop {} {\bibfield
   {journal} {\bibinfo  {journal} {Review of Scientific Instruments}\ }\textbf
  {\bibinfo {volume} {87}},\ \bibinfo {pages} {114903} (\bibinfo {year}
  {2016})}\BibitemShut {NoStop}%
\bibitem [{\citenamefont {Nos\'e}(1984)}]{Nose1984}%
  \BibitemOpen
  \bibfield  {author} {\bibinfo {author} {\bibfnamefont {S.}~\bibnamefont
  {Nos\'e}},\ }\href@noop {} {\bibfield  {journal} {\bibinfo  {journal} {{The
  Journal of Chemical Physics}}\ }\textbf {\bibinfo {volume} {{81}}},\ \bibinfo
  {pages} {{511}} (\bibinfo {year} {{1984}})}\BibitemShut {NoStop}%
\bibitem [{\citenamefont {Baldereschi}(1973)}]{Baldereschi1973}%
  \BibitemOpen
  \bibfield  {author} {\bibinfo {author} {\bibfnamefont {A.}~\bibnamefont
  {Baldereschi}},\ }\href {\doibase {10.1103/PhysRevB.7.5212}} {\bibfield
  {journal} {\bibinfo  {journal} {{Physical Review B}}\ }\textbf {\bibinfo
  {volume} {{7}}},\ \bibinfo {pages} {{5212}} (\bibinfo {year}
  {{1973}})}\BibitemShut {NoStop}%
\bibitem [{\citenamefont {Kubo}(1957)}]{Kubo1957}%
  \BibitemOpen
  \bibfield  {author} {\bibinfo {author} {\bibfnamefont {R.}~\bibnamefont
  {Kubo}},\ }\href@noop {} {\bibfield  {journal} {\bibinfo  {journal} {J. Phys.
  Soc. Jpn.}\ }\textbf {\bibinfo {volume} {12}},\ \bibinfo {pages} {570}
  (\bibinfo {year} {1957})}\BibitemShut {NoStop}%
\bibitem [{\citenamefont {Greenwood}(1958)}]{Greenwood1958}%
  \BibitemOpen
  \bibfield  {author} {\bibinfo {author} {\bibfnamefont {D.~A.}\ \bibnamefont
  {Greenwood}},\ }\href@noop {} {\bibfield  {journal} {\bibinfo  {journal}
  {Proc. Phys. Soc.}\ }\textbf {\bibinfo {volume} {71}},\ \bibinfo {pages}
  {585} (\bibinfo {year} {1958})}\BibitemShut {NoStop}%
\bibitem [{\citenamefont {Holst}\ \emph {et~al.}(2011)\citenamefont {Holst},
  \citenamefont {French},\ and\ \citenamefont {Redmer}}]{Holst2011}%
  \BibitemOpen
  \bibfield  {author} {\bibinfo {author} {\bibfnamefont {B.}~\bibnamefont
  {Holst}}, \bibinfo {author} {\bibfnamefont {M.}~\bibnamefont {French}}, \
  and\ \bibinfo {author} {\bibfnamefont {R.}~\bibnamefont {Redmer}},\
  }\href@noop {} {\bibfield  {journal} {\bibinfo  {journal} {Physical Review
  B}\ }\textbf {\bibinfo {volume} {83}},\ \bibinfo {pages} {235120} (\bibinfo
  {year} {2011})}\BibitemShut {NoStop}%
\bibitem [{\citenamefont {Gajdo\ifmmode~\check{s}\else \v{s}\fi{}}\ \emph
  {et~al.}(2006)\citenamefont {Gajdo\ifmmode~\check{s}\else \v{s}\fi{}},
  \citenamefont {Hummer}, \citenamefont {Kresse}, \citenamefont
  {Furthm\"uller},\ and\ \citenamefont {Bechstedt}}]{Gajdos2006}%
  \BibitemOpen
  \bibfield  {author} {\bibinfo {author} {\bibfnamefont {M.}~\bibnamefont
  {Gajdo\ifmmode~\check{s}\else \v{s}\fi{}}}, \bibinfo {author} {\bibfnamefont
  {K.}~\bibnamefont {Hummer}}, \bibinfo {author} {\bibfnamefont
  {G.}~\bibnamefont {Kresse}}, \bibinfo {author} {\bibfnamefont
  {J.}~\bibnamefont {Furthm\"uller}}, \ and\ \bibinfo {author} {\bibfnamefont
  {F.}~\bibnamefont {Bechstedt}},\ }\href@noop {} {\bibfield  {journal}
  {\bibinfo  {journal} {Phys. Rev. B}\ }\textbf {\bibinfo {volume} {73}},\
  \bibinfo {pages} {045112} (\bibinfo {year} {2006})}\BibitemShut {NoStop}%
\bibitem [{\citenamefont {Berens}\ \emph {et~al.}(1983)\citenamefont {Berens},
  \citenamefont {Mackay}, \citenamefont {White},\ and\ \citenamefont
  {Wilson}}]{Berens1983}%
  \BibitemOpen
  \bibfield  {author} {\bibinfo {author} {\bibfnamefont {P.~H.}\ \bibnamefont
  {Berens}}, \bibinfo {author} {\bibfnamefont {D.~H.~J.}\ \bibnamefont
  {Mackay}}, \bibinfo {author} {\bibfnamefont {G.~M.}\ \bibnamefont {White}}, \
  and\ \bibinfo {author} {\bibfnamefont {K.~R.}\ \bibnamefont {Wilson}},\
  }\href {\doibase DOI:10.1063/1.446044} {\bibfield  {journal} {\bibinfo
  {journal} {The Journal of Chemical Physics}\ }\textbf {\bibinfo {volume}
  {79}},\ \bibinfo {pages} {2375} (\bibinfo {year} {1983})}\BibitemShut
  {NoStop}%
\end{thebibliography}%
\end{document}